\documentclass[letterpaper, 12pt, journal, onecolumn]{IEEEtran}

\usepackage{cite}
\usepackage{amsmath,amssymb,amsfonts}

\usepackage{algorithmic}
\usepackage{graphicx}
\usepackage{textcomp}

\usepackage[usenames,dvipsnames,table]{xcolor}

\usepackage{soul}
\usepackage[normalem]{ulem}

\usepackage{url}
\usepackage{graphicx}
\usepackage{color}
\usepackage{cite}
\usepackage[hidelinks]{hyperref}
\usepackage{amsfonts}
\usepackage{amsmath,multirow} 
\usepackage{algorithm}
\usepackage{mathtools}
\usepackage{caption}
\usepackage{subcaption}

\usepackage{tikz}
\usepackage{pgfplots}
\usepgfplotslibrary{groupplots}

\usepackage{bm}

\newtheorem{theorem}{Theorem}

\newtheorem{assumption}{Assumption}

\newtheorem{lemma}{Lemma}
\newtheorem{remark}{Remark}
\newtheorem{definition}{Definition}

\def\algname/{\texttt{DATA-DRIVEN PRONTO}}
\def\projectionop/{PO}

\newcommand{\R}{{\mathbb{R}}} 
\newcommand{\N}{{\mathbb{N}}} 

\newcommand\oprocendsymbol{\hbox{$\square$}}
\newcommand\oprocend{\relax\ifmmode\else\unskip\hfill\fi\oprocendsymbol}

\DeclareMathOperator*{\subj}{subj. to}

\newcommand{\dimx}{{n}}
\newcommand{\dimu}{{m}}

\newcommand{\dimmu}{{\dimu T}}

\newcommand{\bx}{\bm{x}}
\newcommand{\bu}{\bm{u}}
\newcommand{\bdx}{\bm{d_x}}
\newcommand{\bdu}{\bm{d_u}}
\newcommand{\bDx}{\bm{\Delta X}}
\newcommand{\bDu}{\bm{\Delta U}}
\newcommand{\bDA}{\bm{\Delta A}}
\newcommand{\bDB}{\bm{\Delta B}}

\newcommand{\bdeltax}{\bm{\Delta}\bm{x}}
\newcommand{\bdeltau}{\bm{\Delta}\bm{u}}
\newcommand{\hbdeltax}{\bm{\Delta}\hat{\bm{x}}}
\newcommand{\hbdeltau}{\bm{\Delta}\hat{\bm{u}}}
\newcommand{\deltax}{\Delta x}
\newcommand{\deltau}{\Delta u}
\newcommand{\hdeltax}{\Delta \hat{x}}
\newcommand{\hdeltau}{\Delta \hat{u}}

\newcommand{\balpha}{\bm{\alpha}}
\newcommand{\bmu}{\bm{\mu}}

\DeclareMathOperator*{\argmin}{argmin}
\let\min\relax
\DeclareMathOperator*{\min}{min}
\allowdisplaybreaks

\DeclareMathOperator*{\blk}{diag}
\DeclareMathOperator*{\diag}{diag}

\newcommand{\dynamics}{f}

\newcommand{\stepsize}{\gamma}

\newcommand{\curve}{\xi}
\newcommand{\traj}{\eta}

\newcommand{\dd}{{\nabla}}

\newcommand{\pro}{{\mathcal{P}}}
\newcommand{\iter}{{k}}

\newcommand{\Tr}{{\mathcal{T}}}
\newcommand{\K}{{\mathcal{K}}}
\newcommand{\KL}{{\mathcal{KL}}}

\newcommand{\sx}{n(T+1)}
\newcommand{\su}{mT}

\newcommand{\xinit}{{x_{\text{init}}}}

\newcommand{\map}[3]{#1: #2 \rightarrow #3}

\title{\bf{\algname/: \\
		a Model-free Solution for Numerical Optimal Control}}

\begin{document}

\author{Marco Borghesi, Lorenzo Sforni, Giuseppe Notarstefano
	\thanks{M. Borghesi, L. Sforni, and G. Notarstefano are with the Department of Electrical, Electronic, and Information Engineering, Alma Mater Studiorum - Universit\`a di Bologna, 40136 Bologna, Italy. The corresponding author is M.~Borghesi (e-mail: m.borghesi@unibo.it).}
}

\maketitle

\begin{abstract}
   This article addresses the problem of data-driven numerical optimal
   control for unknown nonlinear systems. In our scenario, we suppose
   to have the possibility of performing multiple experiments (or
   simulations) on the system. Experiments are performed by relying on
   a data-driven tracking controller able to steer the system towards a
   desired reference. Our proposed \algname/ algorithm iteratively
   refines a tentative solution of the optimal control problem by
   computing an approximate descent direction via a local trajectory
   perturbation. At each iteration, multiple trajectories are gathered
   by perturbing the current trajectory with a suitable dither signal,
   and then used to obtain a data-driven, time-varying linearization. The
   exploration is guided by the tracking controller, so that perturbed
   trajectories are obtained in closed loop. We show local convergence of
   \algname/ to a ball about an isolated optimal solution, whose
   radius depends on the amplitude of the dither signal.
   We corroborate the theoretical results by applying it to an
   underactuated robot.
\end{abstract}


\section{Introduction}\label{sec:intro}
Nonlinear optimal control
problems are prevalent in various applications within the fields of Automation and Robotics, where the goal is to develop a control strategy that, when applied to a dynamical system, minimizes a specified performance index~\cite{diehl2011numerical, bryson2018applied, nocedal1999numerical, bertsekas1997nonlinear}.
In general, the solution of Non-Linear Optimal Control Problems (NL-OCPs) strongly relies on a valid model of the system under control which, if inaccurate, can lead to the design of trajectories which are suboptimal for the true system.
In this paper, we propose a combination of identification techniques together with 
optimal control which can solve a NL-OCP in a model-free setting.
In particular, we focus on solving a finite-horizon NL-OCP where the dynamics is unknown but there is the possibility of performing multiple deterministic experiments (or simulations) from the same initial condition. 
Retrieving an optimal trajectory for this kind of setups is important, for example, in several industrial systems, where optimized execution of a repeated task results in significant savings. 
Pioneers in this field were the so called \textit{Repetitive Control} (RC) and \textit{Iterative Learning Control} (ILC)~\cite{bristow2006survey, ahn2007iterative}
by means of multiple repetitions of that specific tracking problem.
Recent works address the problem of including a cost function in the ILC framework and provide suboptimality guarantees for this type of approach,\cite{baumgartner2020zero, baumgartner2023local}. 
Besides RC and ILC, we provide an overview of two fields that deal with the problem of data-driven optimal control, distinguishing between the Reinforcement Learning (RL) and Data-driven optimization approaches (where, since the two fields have lots in common, we stress the fact that this distinction is done arbitrarily, and several works may be placed in both categories).

\subsubsection*{Reinforcement Learning}

Rooted in the principles of dynamic programming \cite{bellman2015applied}
, the field of RL addresses the problem of learning a control policy that minimizes (maximizes) a cumulative cost (reward) through interaction with the environment \cite{sutton2018reinforcement}. 
Learning occurs, in general, during a dedicated training phase, where the agent explores the environment and progressively associates \emph{good} actions with rewards and \emph{bad} actions with penalties.
An early reference in this field is \cite{de2003linear}, where the authors propose to solve the optimal control problem by imposing a linear parametrization of the value function, and finding its parameters via a linear program constrained by collected data samples. 
In \cite{bottcher2022ai} and references therein, Neural Networks (NN) are employed to parametrize the control policy, and gradient descent is used to find the optimal network weights.
In \cite{ma2023reinforcement}, the authors demonstrate through a practical application how combining a classical technique like ILC with RL can significantly reduce the amount of data required for the training.
More recently, \cite{eberhard2024pontryagin} explores the novel idea of using RL to obtain the open-loop optimal input, rather than deriving a feedback control policy.
The challenge of guaranteeing the safety during the training phase is addressed, in general, by including a controller which leverages approximate system knowledge \cite{fisac2018general, zanon2020safe, brunke2022safe} to guarantee constraint satisfaction. 
A research field strictly related to data-driven optimal control is model predictive control,
in which data-driven methods take into account the fact that the system \cite{saviolo2023active, coulson2019data, berberich2020data, hewing2020learning}, or some of the constraints \cite{batkovic2022safe} may be unknown.

\subsubsection*{Data-driven optimization}

Parallel to the RL paradigm, numerical optimization techniques capable of handling 
problems with parametric uncertainties were developed throught the years.
Extremum seeking, black-box optimization, derivative-free optimization and simulation optimization \cite{ariyur2003real, alarie2021two, conn2009introduction, rios2013derivative, amaran2016simulation} are techniques that solve an optimization problem when the explicit cost function, its derivatives, or the constraints are not available to the optimization process. In general, in these cases the idea is to substitute the missing knowledge by cleverly probing the cost function, possibly exploiting the knowledge of some known property (for example, in \cite{sabug2020use, sabug2021smgo, galbiati2022direct} the authors leverage on Lipschitz continuity). 
These methods require in general a high number of experiments, since they do not exploit the particular structure of NL-OCPs, which has been instead studied by several model-based iterative and efficient algorithms  
\cite{diehl2011numerical, bryson2018applied, nocedal1999numerical, bertsekas1997nonlinear}. 
This peculiar structure is leveraged in \cite{chen2019hardware}, where the authors propose an algorithm to iteratively solve a NL-OCP requiring only partial knowledge of the dynamics. 
We refer the reader to \cite{prag2022toward, de2023learning, recht2019tour} for other ways of solving in a data-driven way NL-OCPs.

\subsubsection*{Article Contribution}

The main contribution given in this paper is \algname/, an algorithm for nonlinear optimal control extending the applicability of \texttt{PRONTO} \cite{hauser2002projection} 
to the model-free scenario.
In particular, \algname/ leverages the possibility to probe the system dynamics in a safe way via the application of an independently-designed control law. 
At each algorithm iteration, we perform $L$ experiments (simulations) in which we use the given controller plus an exploration dither to gather data in the neighbourhood of the current trajectory. These data are then used to identify the dynamics linearizations about the current trajectory, substituting the need for a model in the optimization process.
The proposed algorithm achieves the following desirable features: 
i) avoiding parametrization errors: \algname/ uses data to identify local first-order approximations of the dynamics, thus avoiding errors due to a global, inexact parametrization of the nonlinear dynamics or optimal policy;
ii) being model-free: \algname/ itself does not require any knowledge of the system; however, partial system knowledge can be 
used to enhance the numerical stability of the algorithm, e.g., improving the design of a closed-loop control law; 
iii) being data-efficient: thanks to the efficient algorithm structure, the number of required experiments (simulations) needed is comparable, at each algorithm iteration, with the system dimension.
We provide theoretical guarantees on the algorithm convergence properties, together with insights on how to choose its parameters. The proposed algorithm is proved to be locally uniformly ultimately bounded to a ball about the optimal solution whose radius depends on the amplitude of the dither used to achieve successful identification.

This article is organized as follows. In Section~\ref{sec:setup}, we introduce the considered nonlinear optimal control problem and we give preliminaries on \texttt{PRONTO}.
In Section~\ref{sec:ddpronto} we present \algname/. Theoretical guarantees on its convergence are given, and its parameters are discussed.
In Section~\ref{sec:analysis}, we analyze the algorithm convergence via intermediate results, and the section concludes with the proof of the main theorem.
In Section~\ref{sec:simulations}, we show with a numerical example the effectiveness of the proposed algorithm, while Section~\ref{sec:conclusions} concludes the article. The proofs are in Appendix.

\subsubsection*{Notation}\label{sec:notation}
Given two vectors $a \in \R^n, b \in \R^m$, we denote as $\left( a, b \right)$ the column vector stacking $a$ and $b$. Given square matrices $A_0 \in \R^{a_0\times a_0}, \ldots, A_n \in \R^{a_n\times a_n}$, $A=\blk (A_0, \ldots, A_n)$ is the block diagonal matrix storing on its diagonal all the matrices $A_i$ and zeros in all other elements.
We denote stacks of vectors with bold letters, e.g., given $x_0, x_1, \ldots, x_T \in \R^n$, we write $\bx$ to denote $\bx=\left(x_0, x_1, \ldots x_{T}\right) \in \R^{\sx}$. 
A $C^n$ function is a function which is $n$-times continuously differentiable. 
Consider a smooth function $f: \R^n \times \R^m \rightarrow \R^m$. 
We define as $\dd_1 f (\bar{x}, \bar{u}) \in \R^{n\times m}$ the transpose of the Jacobian of $f$ with respect to the first argument evaluated at $\bar{x}, \bar{u}$.
We denote as $\mathbb{B}_p(\bar{x})\subset \R^n$ the open ball of radius $p$ centered in $\bar{x} \in \R^n$;
in case $\bar{x}=0$, we omit the argument and write just $\mathbb{B}_p$ (and similarly, we use $\bm{\text{B}}_p(\bar{x})$ when the ball is closed).
When dealing with vectors we denote as $\|\cdot \|$ the Euclidian norm. 
When dealing with matrices, $\|\cdot \|$ is the Spectral norm.
A class $\K$ function is a function $\alpha: \R_{\geq 0} \rightarrow \R_{\geq 0}$ such that it is strictly increasing and $\alpha(0)=0$.
A class $\KL$ function is a function $\phi: [0, a) \times \R_{\geq 0} \rightarrow \R_{\geq 0}$ such that $\phi(r, s)$ is a class $\K$ function for all fixed $s\in \R_{\geq 0}$, and $\phi(r, s)$, fixed $r$, is decreasing and such that $\phi(r, s)\rightarrow 0$ for $s\rightarrow \infty$.
The condition number of a matrix $M\in\R^{n\times n}$ is denoted as $\kappa(M)$.
Finally, a function $o_x(y)$ is a function of $x, y$ such that $\lim_{y\to 0} \|o_x(y)\|\|y\|^{-1}=0$.

\section{Problem Setup and Preliminaries}\label{sec:setup}

In this first section, we introduce the problem setup together with some preliminaries on model-based optimal control of nonlinear systems.

\subsection{Model-free Nonlinear Optimal Control}

In this article, we consider nonlinear systems described by the discrete-time dynamics 
\begin{equation}\label{eq:dynamics}
x_{t+1}=f(x_t, u_t), \qquad x_0 = \xinit 
\end{equation}
where $\map{\dynamics}{\R^\dimx \times \R^\dimu}{\R^\dimx}$ is the dynamics and $x_t \in\R^\dimx,$ $u_t\in \R^\dimu$ are, respectively, the state of the system and the control input at time $t \in \N$. The initial condition is fixed to be $\xinit \in \R^\dimx$.
We assume dynamics~\eqref{eq:dynamics} to be \emph{unknown}, i.e., we do not have access to an explicit form of $f$.
Nevertheless, we assume to be able to actuate the system with given input sequence $u_0, \ldots, u_{T-1}$ and to measure the noiseless states $x_0, \ldots, x_{T}$. 
Following the notation introduced in Section~\ref{sec:notation}, we define those sequences as
\begin{equation}\label{eq:def_bu_bx}
\begin{split}
\bu &\coloneqq (u_0, \ldots, u_{T-1}) \in \R^{\dimmu}\\
\bx &\coloneqq (x_0, x_1,\ldots, x_T) \in \R^{\dimx (T+1)}.
\end{split}
\end{equation} 
More formally, we assume we can measure \emph{trajectories} of system~\eqref{eq:dynamics}, i.e.,
state-input sequences satisfying the following definition.
\begin{definition}[System trajectory]\label{def:trajectory}
The pair $(\bx, \bu)$ 
is a trajectory of system~\eqref{eq:dynamics} if it satisfies
\begin{equation}
x_{t+1}=f(x_t, u_t)
\end{equation}
for all $t=0,\ldots,T-1$ with $x_0=\xinit$. 
\end{definition}
Compactly, we denote a trajectory as $\traj \coloneqq (\bx,\bu) \in \R^s$, where $s = s_x + s_u$ and $s_x = \sx, s_u = \su$.
\begin{definition}[Trajectory manifold]\label{def:traj_manifold}
We denote as $\Tr\subset\R^s$ the set of all the trajectories of~\eqref{eq:dynamics} as per Definition~\ref{def:trajectory} of fixed length $T$ and initial condition $x_{\text{init}}$.
Notice that, by defining $h:\R^s\to \R^{s_x}$,
\begin{equation}\label{eq:dyn_constrint}
h(\bx, \bu) = 
\begin{bmatrix}
	x_0 - \xinit\\
	\vdots\\
	x_{T}-f(x_{T-1}, u_{T-1})
\end{bmatrix},
\end{equation}
it holds that $(\bx, \bu)\in \Tr\iff h(\bx, \bu)=0$. 
\end{definition}
Conversely, a generic element of $\R^s$ not necessarily satisfying Definition~\ref{def:trajectory} is said to be a \emph{curve} and it is denoted as $\curve \coloneqq (\balpha, \bmu) \in \R^{s}$, with
\begin{align}
\label{eq:def_balpha_bmu}
\begin{split}
\balpha &\coloneqq (\alpha_0, \ldots, \alpha_{T}) \in \R^{s_x},
\\
\bmu &\coloneqq (\mu_0, \ldots, \mu_{T-1}) \in \R^{s_u}.
\end{split}
\end{align}
Our objective is to design input sequences $\bu$ for the unknown system~\eqref{eq:dynamics}
such that the resulting trajectory $(\bx,\bu)$ minimizes a nonlinear performance index 
\begin{align}\label{eq:cost}
\ell(\bx, \bu) \coloneqq \sum_{t=0}^{T-1}\ell_t(x_t, u_t) + \ell_T(x_T),
\end{align}
where $\map{\ell_t}{\R^\dimx \times \R^\dimu}{\R_{\ge 0}}$ is the so-called stage cost and $\map{\ell_T}{\R^\dimx}{\R_{\ge 0}}$ is the terminal cost.
Compactly, we aim at solving the following optimal control problem
\begin{align}\label{prob:optcon_problem}
\min_{\bm{x}, \bm{u}}\;\;\; & \sum_{t=0}^{T-1}\ell_t(x_t, u_t) + \ell_T(x_T)\\
\subj \;\;\;& x_{t+1} = f(x_t, u_t), \;\;\; t = 0, \ldots, T-1,\nonumber\\
& x_0 = \xinit.\nonumber
\end{align}
We introduce the following hypothesis on the smoothness of the dynamics and the cost function.
\begin{assumption}[Smooth Cost and Dynamics]\label{hp:f_ell_regularity}
The dynamics $f$ in~\eqref{eq:dynamics} and the cost function $\ell$ in~\eqref{eq:cost} are twice continuously differentiable in their arguments, i.e., they belong to class $C^2$.
\end{assumption}
As it will be clearer further in the paper, 
we also introduce the following assumptions regarding the second-order derivatives of the cost function~\eqref{eq:cost}.
\begin{assumption}[Positive Definite Cost Hessian]\label{hp:newton}
For all state and input sequences $(\bx, \bu)\in \R^s$, it holds that $\dd^2\ell(\bx, \bu)>0$.
\end{assumption}
Assumption \ref{hp:newton} guarantees the possibility 
of finding a valid descent direction. 
Furthermore, since the cost $\ell$ is tipically user-defined, it is possible to readily satisfy Assumption~\ref{hp:newton}.
We highlight that, in this paper, the key challenge is that dynamics~\eqref{eq:dynamics} and its derivatives are not accessible in their explicit form, i.e., they are not available to the solver. 

\subsection{A model-based algorithm: \texttt{PRONTO} and the Projection Operator approach}
\label{sec:pronto}
In this section we present a discrete-time version of the optimal control algorithm \texttt{PRONTO} (on which we build \algname/), 
that has been proposed in~\cite{hauser2002projection} for the continuous-time framework. 
The underlying idea is to leverage a feedback policy mapping generic elements $(\balpha, \bmu)$ of the space $\R^s$, the so-called curves, into the set $\Tr$ of trajectories feasible for the dynamics~\eqref{eq:dynamics}.
This feedback policy is assumed to be implemented via the nonlinear tracking system
\begin{align}\label{eq:closed_loop_system}
u_t &= \pi(\alpha_t, \mu_t, x_t, t)\\
x_{t+1} &= f(x_t, u_t), \qquad x_0 =\xinit\nonumber,
\end{align}
for all $t=0, \ldots, T-1$, 
where $\map{\pi}{\R^\dimx \times \R^\dimu \times \R^\dimx \times \R}{\R^\dimu}$ is a generic tracking controller.
The interconnection~\eqref{eq:closed_loop_system} between the control law $\pi$ and the dynamics $f$ can be seen as a \emph{projection operator} which projects generic curves $(\balpha, \bmu)$ onto the trajectory manifold of system \eqref{eq:dynamics}, i.e., it implements a map $(\balpha, \bmu)\mapsto (\bx, \bu)$.
We refer to~\cite{hauser2002projection} for an in-depth discussion.
More in detail, for all iterations indexed by $\iter$, 
the methodology proposed in~\cite{hauser2002projection}
seeks for an update direction $(\bdeltax^\iter, \bdeltau^\iter)$ onto the tangent space of the trajectory manifold at the current solution trajectory $(\bx^\iter, \bu^\iter)$ (\textbf{Step $\boldsymbol{1}$}  of Algorithm \ref{alg:PRONTO}).
The descent direction is obtained by solving LQR problem \eqref{eq:pronto_descent},
where $A_t^\iter, B_t^\iter$, $q_t^\iter, r_t^\iter$ are defined as
\begin{equation}\label{eq:dyn_linearizations}
\begin{split}
&A_t^\iter \coloneqq \dd_1 f(x_t^\iter, u_t^\iter)^\top, \hspace{1cm} B_t^\iter \coloneqq\dd_2 f (x_t^\iter, u_t^\iter)^\top, \\
&q_t^\iter \coloneqq \dd_{1} \ell_t(x_t^\iter, u_t^\iter)^\top, \hspace{1cm} r_t^\iter \coloneqq \dd_{2} \ell_t(x_t^\iter, u_t^\iter)^\top,
\end{split}
\end{equation}
and $Q_t^\iter, Q_T^\iter \in \R^{\dimx \times \dimx}$, $S_t^\iter \in \R^{\dimx \times \dimu}$ and $R_t^\iter \in \R^{\dimu \times \dimu}$ are defined as 
\begin{equation}\label{eq:simplified_pronto}
\begin{split}
Q_t^\iter &\coloneqq \dd^{2}_{11}\ell_t(x_t^\iter, u_t^\iter),  \hspace{1cm}
S_t^\iter \coloneqq \dd^{2}_{12}\ell_t(x_t^\iter, u_t^\iter),\\
R_t^\iter &\coloneqq \dd^{2}_{22}\ell_t(x_t^\iter, u_t^\iter),  \hspace{1cm}
Q_T^\iter \coloneqq \dd^{2}_{11}\ell_t(x_T^\iter).
\end{split}
\end{equation}

After computing the descent direction, \texttt{PRONTO} updates the estimate of the solution $(\balpha^{\iter+1}, \bmu^{\iter+1})$ according to
\begin{align}
\begin{bmatrix}
\balpha^{\iter+1}\\
\bmu^{\iter+1}
\end{bmatrix}
=
\begin{bmatrix}
\bx^\iter\\
\bmu^\iter
\end{bmatrix}
+\stepsize
\begin{bmatrix}
\bdeltax^\iter\\
\bdeltau^\iter
\end{bmatrix},		
\end{align}
where $\stepsize\in(0,1]$ is an appropriate stepsize (\textbf{Step $\boldsymbol{2}$}  of Algorithm \ref{alg:PRONTO}), 
and the updated trajectory $(\bx^{\iter+1}, \bu^{\iter+1})$ is then obtained from $(\balpha^{\iter+1}, \bmu^{\iter + 1})$ via~\eqref{eq:closed_loop_system}  (\textbf{Step $\boldsymbol{3}$}  of Algorithm \ref{alg:PRONTO}). 
\begin{remark}[Approximations of descent step]\label{remark:approximation}
In \eqref{eq:simplified_pronto}, we consider a quadratic part of the cost which is positive definite by Assumption \ref{hp:newton}.
However, in order not to lose convergence, it is sufficient to choose any positive definite matrix (e.g., using the identity matrix or regularizing the cost)\cite[Cor. 4.3]{diehl2011numerical}. 
\end{remark}

Algorithm~\ref{alg:PRONTO} recaps the procedure described so far.
\begin{algorithm}[]
\begin{algorithmic}[0]
\caption{\texttt{PRONTO}}\label{alg:PRONTO}
\STATE \hspace{-0cm}\textbf{Initialization:} initial trajectory $(\bx^0, \bu^0)$.
\FOR{$\iter = 0, 1, 2 \ldots$}
\STATE \textbf{Step $\bm{1}$:} find descent direction $(\bdeltax^\iter,\bdeltau^\iter)$ by solving
\begin{align}\label{eq:pronto_descent}
	\min_{\bdeltax,\bdeltau} \;\; & \!\!\sum_{t=0}^{T-1} 
	\frac{1}{2}
	\hspace{-0.05cm}
	\begin{bmatrix}
		\deltax_t \\ \deltau_t 
	\end{bmatrix}^{\mspace{-6mu}\top}
	\hspace{-0.15cm}
	\begin{bmatrix}
		Q_t^\iter & S^\iter_t \\
		S^{\top, \iter}_t & R^\iter_t
	\end{bmatrix}
	\hspace{-0.15cm}
	\begin{bmatrix}
		\deltax_t \\ \deltau_t 
	\end{bmatrix}
	\!\!+\!
	\begin{bmatrix}
		q^\iter_t\\
		r^\iter_t
	\end{bmatrix}^{\mspace{-6mu}\top} 
	\hspace{-0.15cm}
	\begin{bmatrix}
		\deltax_t \\ \deltau_t
	\end{bmatrix} 
	\hspace{-0.05cm}
	\nonumber\\
	& \hspace{1cm}
	+  \frac{1}{2}\deltax_T^\top
	Q_T^\iter \deltax_T + q_T^{\iter\top} \deltax_T 
	\nonumber\\
	\subj \: & \: \deltax_{t+1} = A_t^\iter \deltax_t + B_t^\iter \deltau_t,\\
	& \deltax_{0} = 0,  \;\; t = 0, \ldots, T-1.
	\nonumber
\end{align}
\STATE \textbf{Step $\bm{2}$:} update curve $(\balpha^{\iter+1}, \bmu^{\iter+1})$:
\begin{equation}\label{alg:PRONTO:update_curve}
	\begin{split}
		\alpha^{\iter + 1}_t & = x^\iter_t + \stepsize \deltax^\iter_t
		\\
		\mu^{\iter + 1}_t & = u^\iter_t + \stepsize \deltau^\iter_t,
	\end{split}
\end{equation}
with $t=0, \ldots, T-1$.
\vspace{0.1cm}
\STATE \textbf{Step $\bm{3}$:} find new trajectory $(\bx^{\iter+1}, \bu^{\iter+1})$:
\STATE
\vspace{-0.4cm}
\begin{equation}
	\begin{split}
		u^{\iter+1}_t &= \pi(\alpha^{\iter+1}_t, \mu^{\iter+1}_t, x^{\iter+1}_t, t)\\
		x^{\iter+1}_{t+1} &= f(x^{\iter+1}_t, u^{\iter+1}_t), \;\;\; x^{\iter+1}_0 = \xinit,
	\end{split}
\end{equation}
with $t=0, \ldots, T-1$.
\ENDFOR
\end{algorithmic}
\end{algorithm}

\section{Data-driven PRONTO}\label{sec:ddpronto}

We are now ready to present our data-driven
optimal control algorithm \algname/, whose pseudocode is summarized in
Algorithm~\ref{alg:DD-PRONTO}.
\begin{algorithm}
\begin{algorithmic}[0]
\caption{\texttt{DATA-DRIVEN PRONTO}}
\label{alg:DD-PRONTO}
\STATE \hspace{-0cm}\textbf{Initialization:} initial trajectory $(\bx^0, \bu^0)$.
\FOR{$\iter = 0, 1, 2 \ldots$}
\STATE \textbf{\emph{Learning}}\\
\STATE \textbf{Step $\bm{L1}$:} gather $L$ trajectories perturbation of $(\bm{x}^\iter, \bm{u}^\iter)$ via closed-loop experiments
\begin{align}\label{eq:closed_loop_experiment}
	\begin{split}
		\hat{u}^{i, \iter}_t &= \pi(x_t^\iter + d^{i, \iter}_{x, t}, u_t^\iter + d^{i, \iter}_{u, t}, \hat{x}^{i, \iter}_t, t)\\
		\hat{x}^{i, \iter}_{t+1} &= f(\hat{x}^{i, \iter}_t, \hat{u}^{i, \iter}_t), \qquad \hat{x}^{i, \iter}_0 = \xinit + d_{x, 0}^{i, \iter}.
	\end{split}
\end{align}
Build $\Delta X_t^\iter, 	\Delta U_t^\iter, \Delta X^{+,\iter}_t$ for $t=0, \ldots T-1$ as in~\eqref{eq:batches}.
\STATE \textbf{Step $\bm{L2}$:} for all $t=0, \ldots, T-1$, estimate the linearizations of the dynamics:
\begin{align}\label{eq:identification}
	\begin{bmatrix}
		\hat{A}^\iter_t & \hat{B}^\iter_t
	\end{bmatrix}
	= \Delta X_t^{+,\iter}
	\begin{bmatrix}
		\Delta X^\iter_t \\
		\Delta U^\iter_t
	\end{bmatrix}^\dagger,
\end{align}
with $\Delta X_t^\iter, \Delta U_t^\iter, \Delta X_t^{+, \iter}$ in~\eqref{eq:batches}.
\STATE \textbf{\emph{Optimization}}
\STATE \textbf{Step $\bm{O1}$:} solve the approximate problem
\begin{align}\label{prob:pronto_approx}
	\min_{\hbdeltax,\hbdeltau} \;\; & \!\!\sum_{t=0}^{T-1} 
	\frac{1}{2}
	\begin{bmatrix}
		\Delta \hat{x}_t\\
		\Delta \hat{u}_t
	\end{bmatrix}^{\mspace{-6mu}\top}
	\hspace{-0.15cm}
	\begin{bmatrix}
		Q^\iter_t & S_t^\iter \\
		S^{\top, \iter}_t & R^\iter_t
	\end{bmatrix}
	\hspace{-0.15cm}
	\begin{bmatrix}
		\Delta \hat{x}_t\\
		\Delta \hat{u}_t
	\end{bmatrix}
	\!\!
	+
	\!
	\begin{bmatrix}
		q^\iter_t\\
		r^\iter_t
	\end{bmatrix}^{\mspace{-6mu}\top} 
	\hspace{-0.15cm}
	\begin{bmatrix}
		\Delta \hat{x}_t\\
		\Delta \hat{u}_t
	\end{bmatrix}\nonumber \\
	& + \frac{1}{2}\Delta \hat{x}_T^\top Q^\iter_T \Delta \hat{x}_T + q_T^{\iter\top} \Delta \hat{x}_T \nonumber \\
	\subj \:  & \: \Delta \hat{x}_{t+1}=\hat{A}^\iter_t\Delta \hat{x}_t + \hat{B}^\iter_t\Delta \hat{u}_t, \\
	& \Delta \hat{x}_0 = 0, \;\;\; t=0, \ldots T-1.
	\nonumber
\end{align}
\STATE \textbf{Step $\bm{O2}$:} update curve $(\balpha^{\iter+1}, \bmu^{\iter+1})$
\begin{equation}
	\begin{split}
		\alpha_t^{\iter+1} & = x_t^\iter + \gamma \hdeltax_t^\iter
		\\
		\mu_t^{\iter+1} & = u_t^\iter +  \gamma \hdeltau_t^\iter,
	\end{split}
\end{equation}
with $t=0, \ldots, T-1$.
\STATE \textbf{Step $\bm{O3}$:} obtain new trajectory $(\bx^{\iter+1}, \bu^{\iter+1})$
\begin{equation}\label{eq:projection_step}
	\begin{split}
		u_t^{\iter+1} & = \pi(\alpha_t^{\iter+1}, \mu_t^{\iter+1}, x_t^{\iter+1},  t)\\
		x_{t+1}^{\iter+1} & = \dynamics (x_t^{\iter+1}, u_t^{\iter+1}),\;\;\; x_0^{\iter+1}=x_\text{init},
	\end{split}
\end{equation}
with $t=0, \ldots, T-1$.
\ENDFOR
\end{algorithmic}
\end{algorithm}

\subsection{The algorithm}\label{subsec:thealgorithm}
Here, we describe how \algname/ is capable of generating a solution to
problem~\eqref{prob:optcon_problem} in a model-free framework by
leveraging on successive \emph{learning} and \emph{optimization}
steps, which are denoted in Algorithm~\ref{alg:DD-PRONTO} by prefixes
$\boldsymbol{L}$ and $\boldsymbol{O}$, respectively.

\paragraph*{\textbf{Step} $\boldsymbol{L1}$: Closed-loop data collection}
At each iteration $\iter$, a set of state-input data is collected in the neighbourhood of the \emph{current} trajectory $(\bx^\iter, \bu^\iter)$ by successive experimental sessions (or simulations) with additional exploration noise.
The collected data are then used to estimate the Jacobians of the unknown dynamics~\eqref{eq:dynamics}. 
More in detail, the $i^{th}$ perturbation of the nominal trajectory $(\bx^\iter, \bu^\iter)$, denoted as  $(\hat{\bx}^{i, \iter}, \hat{\bu}^{i, \iter})$, 
is obtained from the real system by integrating the closed-loop dynamics \eqref{eq:closed_loop_experiment}, 
where $d^{i, \iter}_{x, t} \in \R^n, d^{i, \iter}_{u,t} \in \R^m$ are appropriate exploration dithers injected to guarantee a successful identification. 
We introduce now two hypotheses to characterize the control law $\pi$ and the closed-loop experiments~\eqref{eq:closed_loop_experiment}.
\begin{assumption}[Properties of $\pi$]\label{hp:pi_regularity}
The state-feedback control law $\pi(\alpha, \mu, x, t)$ is twice continuously differentiable in its arguments, i.e., it is $C^2$ and it is designed such that $\pi(\alpha, \mu, \alpha, t) = \mu$ holds for all $\alpha \in \R^n, \mu \in \R^n, t \in \N$.
\end{assumption}
\begin{remark}
This assumption implies that when the reference curve $(\balpha, \bmu)$  is a trajectory of the dynamics~\eqref{eq:dynamics} as per Definition \ref{def:trajectory},
the closed-loop trajectory $(\bx, \bu)$ resulting from \eqref{eq:closed_loop_system} is such that $(\balpha, \bmu)=(\bx, \bu)$.
\end{remark}

\paragraph*{\textbf{Step} $\boldsymbol{L2}$: LTV dynamics identification}
For all $i = 1,\ldots, L$, the perturbations $(\hat{\bx}^{i, \iter}, \hat{\bu}^{i, \iter})$ obtained via~\eqref{eq:closed_loop_experiment}
are used to 
build matrices $\Delta X^\iter_t, \Delta U^\iter_t, \Delta X^{+,\iter}_t$, which are data batches stacking the differences between all the perturbations and the current trajectory, namely
\begin{align}
\Delta X^\iter_t&= [\hat{x}_{t}^{1, \iter}-x^\iter_t, \ldots, \hat{x}_{t}^{L, \iter}-x^\iter_t] \in \R^{n\times L}\nonumber\\
\Delta U^\iter_t&= [\hat{u}_{t}^{1, \iter}-u^\iter_t,\ldots, \hat{u}_{t}^{L, \iter}-u^\iter_t]\in \R^{m \times L}\label{eq:batches}\\
\Delta X^{+,\iter}_t&= [\hat{x}_{t+1}^{1, \iter}-x^\iter_{t+1},\ldots, \hat{x}_{t+1}^{L, \iter}-x^\iter_{t+1}]\in \R^{n\times L}\nonumber.
\end{align}
Data $\Delta X^\iter_t, \Delta U^\iter_t, \Delta X^{+,\iter}_t$ are used to perform a least-squares identification~\eqref{eq:identification} of the matrices $A^\iter_t, B^\iter_t$ for all $t$. 
Denote any isolated local minimum of~\eqref{prob:optcon_problem} as $(\bx^\star, \bu^\star)$. 
The following assumption ensures it is possible to achieve a well-posed identification step in a neighbourhood of the optimal solution.
\begin{assumption}[Well-posed identification]\label{hp:well_posed}
There exists $c>0$ such that, for all dither bounds
$\delta_x, \delta_u>0$ and trajectories
$(\bx, \bu) \in \mathbb{B}_c(\bx^\star,
\bu^\star)\cap \Tr$, there exist $L\geq n+m$ dither
sequences
$d_{x, 0}^i, \ldots, d_{x, {T-1}}^i\in \R^n, d_{u,
0}^i, \ldots, d_{u, {T-1}}^i\in \R^m$,
$i = 1, \ldots, L$, such that
\begin{itemize}
\item[i)] for all $t=0, \ldots, T-1$ the dithers are bounded by
\begin{equation}
	\|d_{x, t}^{i}\| \leq \delta_x  \hspace{2cm} \|d_{u, t}^{i} \| \leq \delta_u
\end{equation}
\item[ii)] the data batches $\Delta X_t, \Delta U_t$ obtained as in \eqref{eq:batches} satisfy, for some $M>0$ and for all $t=0, \ldots, T-1$
\begin{equation}\label{eq:full_rank_batches}
	\kappa \left(
	\begin{bmatrix}
		\Delta X_t\\
		\Delta U_t
	\end{bmatrix}
	\begin{bmatrix}
		\Delta X_t\\
		\Delta U_t
	\end{bmatrix}^\top\right) \leq M,
\end{equation}
\end{itemize}
where $\kappa(\cdot)$ is the condition number of a square matrix.
\end{assumption}

\begin{remark}[On well-conditioning]
Although it is challenging to establish sufficient conditions on the exploration noise to guarantee Assumption~\ref{hp:well_posed} for generic nonlinear dynamics, the proposed strategy allows for the collection of an arbitrary number of trajectory perturbations and the verification of whether the conditions are met using the gathered data.
Furthermore, notice that this assumption is well-characterized for linear time-invariant systems \cite{willems2005note, borghesi2025sufficient}. 
\end{remark}
\vspace{0.1cm}

\paragraph*{\textbf{Step} $\boldsymbol{O1}$: Data-driven descent calculation}
Finally, in Algorithm~\ref{alg:DD-PRONTO} the descent direction is obtained by solving the data-based problem~\eqref{prob:pronto_approx} where the 
linearization of the dynamics is replaced by its estimation based on data~\eqref{eq:batches}.

\noindent
Steps $\boldsymbol{O2}$ and $\boldsymbol{O3}$ are left unchanged from Algorithm~\ref{alg:PRONTO}.

\subsection{Main result}\label{subsec:mainresult}
In the following we present the main result of this paper.
For the sake of readability, for all iteration $\iter$ we denote the solution estimate provided by Algorithm~\ref{alg:DD-PRONTO} as $\traj^\iter = (\bx^\iter, \bu^\iter)$ while each isolated local minimum solution of~\eqref{prob:optcon_problem} is denoted as $\traj^\star = (\bx^\star, \bu^\star)$. 

\begin{theorem}\label{theo:main_result}
Consider Algorithm \ref{alg:DD-PRONTO}.
Let Assumptions~\ref{hp:f_ell_regularity},~\ref{hp:newton},~\ref{hp:pi_regularity} and~\ref{hp:well_posed} hold.  
Then, there exist a stepsize $\gamma^\star>0$, dither bounds $\bar{\delta}_x, \bar{\delta}_u >0$, an iteration $N \in \N$, and a radius $\rho>0$ 
such that, if 
$\delta_x \in(0,\bar{\delta}_x), \delta_u \in(0,\bar{\delta}_u)$, $\traj^0 \in  \mathbb{B}_\rho(\traj^\star)\cap \Tr$, $\gamma \in (0, \gamma^\star]$ and the dither signals are chosen at iteration as per \eqref{eq:full_rank_batches} in Assumption~\ref{hp:well_posed},
then
\begin{align}
	\|\traj^\iter -\traj^\star\| \leq b(\delta_x, \delta_u) &&& \forall k\geq N,
\end{align}
with $b(\delta_x, \delta_u)$ strictly increasing in $\delta_x, \delta_u$ and zero in zero.
\end{theorem}

\vspace{0.2cm}
\noindent
Notice that \algname/ does not identify a specific model of the dynamics, but identifies linearizations of the dynamics about specific trajectories. This avoids the introduction of possible parametrization errors when performing the identification.

\section{Algorithm Analysis}\label{sec:analysis}

We approach the proof of the main theorem in two steps. 
First, we show that 
the solution update strategy implemented by Algorithm~\ref{alg:DD-PRONTO} can be viewed as a perturbed version of Algorithm~\ref{alg:PRONTO} (where the perturbation is introduced by the Jacobian estimation).
Under the assumption that the difference between the descent directions for the full-knowledge problem \eqref{eq:pronto_descent} and the estimated problem~\eqref{prob:pronto_approx} is sufficiently small, we can ensure convergence to a neighborhood of the solution $\eta^\star$.
Second, we demonstrate how to pick the algorithm parameters $\delta_x$ and $\delta_u$ to ensure that the difference between the descent direction obtained by solving problems~\eqref{eq:pronto_descent} and~\eqref{prob:pronto_approx} can be made arbitrarily small.
For the sake of clarity, we denote the current (unperturbed) trajectory at each iteration as $\traj^\iter=(\bx^\iter, \bu^\iter) \in \Tr$. 
To denote solutions of problems~\eqref{eq:pronto_descent} and~\eqref{prob:pronto_approx}, i.e., the descent direction, we introduce the symbols $\zeta^\iter=(\bdeltax^\iter, \bdeltau^\iter) \in \R^s$ and $\hat{\zeta}^\iter=(\hbdeltax^\iter, \hbdeltau^\iter) \in \R^s$, respectively.

\subsection{Practical stability of \texttt{DATA-DRIVEN PRONTO}}\label{sec:errorpronto}
We now study the convergence properties of \texttt{PRONTO} in case of errors in the descent direction calculation.
At first, we briefly introduce a state-space reformulation of \texttt{PRONTO}. 
Then, we study Algorithm \ref{alg:DD-PRONTO} as a perturbed version of Algorithm~\ref{alg:PRONTO}.
We start by rewriting problem \eqref{prob:optcon_problem} as
\begin{equation}\label{prob:h_formulation}
\begin{split}
\min_{\traj} \;\;\; & \ell(\traj)\\
\text{subj. to}\;\;\; & \traj \in \Tr,
\end{split}
\end{equation} 
where $\ell(\cdot)$ and $\Tr$ are given in \eqref{eq:cost} and Definition \ref{def:traj_manifold}, respectively.
We denote the operator associated to the closed-loop system \eqref{eq:closed_loop_system} as $\pro:\R^s \to \Tr$, since it projects any curve $\curve$ into a trajectory $\traj$ by leveraging the tracking controller $\pi$ and integrating the dynamics.
Further details on this operator and its properties can be found in \cite{hauser1998trajectory}.
We then embed $\pro$ in the cost by defining $g(\curve)\coloneqq \ell(\pro(\curve))$ to obtain the \emph{unconstrained} problem formulation
\begin{equation}\label{prob:g_formulation}
\begin{split}
\min_{\curve \in \R^s} \;\;\; & g(\curve),
\end{split}
\end{equation}
which is shown to have the same isolated minima of~\eqref{prob:h_formulation} \cite[Pag. $2$]{hauser2002projection}.
In Algorithm \ref{alg:PRONTO}, 
problem~\eqref{prob:g_formulation} is then solved via a quasi Newton's method which can be rewritten as the the iterative update
\begin{equation}
\label{eq:approx_pronto_dynamics}
\begin{split}
\zeta^\iter &= \argmin_{\zeta \in T_{\traj^\iter}\Tr} \left(\frac{1}{2} \zeta^\top \dd^2 \ell (\traj^\iter) \zeta + \dd\ell (\traj^\iter)^\top\zeta \right)\\
\traj^{\iter+1}	&= \pro(\traj^\iter + \gamma\zeta^\iter).
\end{split}
\end{equation}
Notice that, in~\eqref{eq:approx_pronto_dynamics}, only the second order derivatives of the cost function $\ell$ are considered, instead of those of the composition $g=\ell \circ \pro$ (cf. Remark~\ref{remark:approximation}).
The following lemma provides stability guarantees of the optimal solution $\traj^\star$ for Algorithm~\ref{alg:PRONTO}.

\begin{lemma}\label{lemma:hessian_approximation_convergence}
Consider the update rule for the optimal solution of problem~\eqref{prob:optcon_problem} given by Algorithm \ref{alg:PRONTO} (\texttt{PRONTO}), namely, the discrete-time system \eqref{eq:approx_pronto_dynamics}.
Let Assumptions~\ref{hp:f_ell_regularity}, ~\ref{hp:newton} and~\ref{hp:pi_regularity} hold. Then, there exists $\gamma^\star>0$ such that, if $\gamma \in (0, \gamma^\star]$, the equilibrium $\traj^\star$ is Locally Exponentially Stable for \eqref{eq:approx_pronto_dynamics}. 
%
\end{lemma}
The proof is provided in Appendix~\ref{proof:lemma:hessian_approximation_convergence}.
Given the stability properties of $\traj^\star$ for the update \eqref{eq:approx_pronto_dynamics}, we are now able to provide theoretical guarantees for a perturbed version of the algorithm.
\begin{lemma}\label{lemma:practical_stability}
Consider an update rule for the optimal solution of problem~\eqref{prob:optcon_problem} given by
\begin{equation}\label{eq:perturbed_pronto_dynamics}
\begin{split}
	\zeta^\iter &= \argmin_{\zeta \in T_{\traj^\iter}\Tr} \left(\frac{1}{2} \zeta^\top \dd^2 \ell (\traj^\iter) \zeta + \dd\ell (\traj^\iter)^\top\zeta \right)\\
	\traj^{\iter+1}	&= \pro\left(\traj^\iter + \gamma (\zeta^\iter+\Delta \zeta^\iter)\right),
\end{split}
\end{equation}
where $\Delta \zeta^\iter\in \R^s$ is a bounded perturbation, namely, $\|\Delta \zeta^\iter\|\leq \delta_\zeta$ for all $\iter \in \N$ and for some $\delta_\zeta>0$.
Let Assumptions~\ref{hp:f_ell_regularity},~\ref{hp:newton} and~\ref{hp:pi_regularity} hold.
Then, there exist $\gamma^\star>0$ and
$\bar{\delta}_\zeta>0$ such that, if
$\delta_\zeta \leq \bar{\delta}_\zeta$ and
$\gamma\in (0,\gamma^\star]$, the evolution of \eqref{eq:perturbed_pronto_dynamics}
is Locally Uniformly Ultimately Bounded,
i.e., there exists $N\in \N ,p>0$, a class $\KL$
function $\phi$ and a class $\K$ function
$b$ such that, if
$\traj^0 \in \mathbb{B}_p(\traj^\star)\cap \Tr$, then
it holds
\begin{equation}\label{eq:KL_bound}
\begin{split}
	\| \traj^\iter-\traj^\star\|&\leq\phi(\|\traj_0-\traj^\star \|, k) \hspace{0.6cm} \forall k<N,\\
	\| \traj^\iter-\traj^\star\|&\leq b(\delta_\zeta) \hspace{2cm} \forall k\geq N.
\end{split}
\end{equation} 
%
\end{lemma}
The proof is provided in Appendix~\ref{proof:lemma:practical_stability}.

\subsection{Data-driven descent error}\label{sec:ddLQR}

In the following, we present formal error bounds about the calculation of the descent direction based solely on input dithers.
Since the results in this section are independent on the specific iteration $\iter$ of the algorithm, we omit the superscript $\iter$ for clarity.  
Additionally, to simplify the notation, we define $\bdx$ and $\bdu$ as the stack of exploration dithers across all time steps and perturbations, i.e.,
\begin{equation}\label{eq:bdx_bdu_def}
\begin{split}
\bdx &\coloneqq (d_{x,0}^1, \ldots, d_{x, T-1}^1, \ldots, d_{x,0}^L, \ldots, d_{x,T-1}^L)\in \R^{nTL}\\
\bdu &\coloneqq (d_{u,0}^1, \ldots, d_{u, T-1}^1, \ldots, d_{u,0}^L, \ldots, d_{u,T-1}^L)\in \R^{mTL}.\\
\vphantom{A}
\end{split}
\end{equation}
\vspace{-0.7cm}

The core idea is to show when the error between the data-driven descent direction $\hat{\zeta}=(\hbdeltax, \hbdeltau)$ solution of~\eqref{prob:pronto_approx} and the exact descent direction $\zeta=(\bdeltax, \bdeltau)$ solution of~\eqref{eq:pronto_descent} is a smooth function of the exploration dithers $\bdx, \bdu$.
Let us denote the stack of data matrices~\eqref{eq:batches} obtained by collecting $L$ perturbed trajectories from~\eqref{eq:closed_loop_experiment} as
\begin{equation}
\label{eq:dxu}
\begin{split}
\bDx &\coloneqq (\Delta X_0, \ldots, \Delta X_{T-1})\in \R^{nTL}\\
\bDu &\coloneqq (\Delta U_0, \ldots, \Delta U_{T-1})\in \R^{mTL}.
\end{split}
\end{equation}
First, we show that, for a given trajectory $\traj$, the data matrices $(\bDx, \bDu)$ in~\eqref{eq:dxu} are a smooth function of the exploration dithers $\bdx, \bdu$, namely we show the function
\begin{align}\label{eq:fcn_dxu}
	\Delta_{XU}: &\Tr \times \R^{nTL} \times \R^{mTL}\to \R^{nT L}\times \R^{mTL}\\
	&\traj, \bdx,  \bdu  \longmapsto \Delta_{XU}(\traj, \bdx, \bdu)\coloneqq(\bDx, \bDu)\nonumber
\end{align}

is smooth in its arguments.
Notice that $\Delta_{XU}(\traj, \bdx, \bdu)$ depends both on the exploration dithers of all $i= 1, \ldots, L$ collected perturbations and on the current trajectory $\traj$, where the dependence on $\traj$ accounts for the closed-loop dynamics~\eqref{eq:closed_loop_experiment}.
\begin{lemma}\label{lemma:bounded_perturbations}
Let Assumptions~\ref{hp:f_ell_regularity} and~\ref{hp:pi_regularity} hold. Then, $\Delta_{XU}$ in \eqref{eq:fcn_dxu} is a $C^1$ function of the trajectory $\traj$ and of the exploration dithers $\bdx, \bdu$. Furthermore, 
\begin{equation}
\Delta_{XU}(\traj, 0, 0)=0
\end{equation}
for all system trajectories $\traj\in \Tr$. 
\end{lemma}
The proof is provided in Appendix~\ref{proof:lemma:bounded_perturbations}.
Let 
\begin{equation}\label{eq:deltaA_deltaB}
\begin{split}
\Delta A_t \coloneqq \hat{A}_t - A_t, \hspace{1cm} \Delta B_t \coloneqq \hat{B}_t - B_t,
\end{split}
\end{equation}
denote the error between the matrices estimated using data via~\eqref{eq:identification} and the exact Jacobians of $f$ about the current trajectory $\traj$ at time instant $t$.
Define their stack, for $t=0, \ldots, T-1$, as
\begin{equation}
\label{eq:bDAB}
\begin{split}
\bDA &\coloneqq (\Delta A_0, \ldots, \Delta A_{T-1})\in \R^{n^2T}\\
\bDB &\coloneqq (\Delta B_0, \ldots, \Delta B_{T-1})\in \R^{nmT}.
\end{split}
\end{equation}
The stack of estimation errors in~\eqref{eq:bDAB} can be written as a function of both the current trajectory $\traj$ and the data matrices $\bDx, \bDu$, which define the exact Jacobians and their data-based estimation, respectively.
We denote this function as

\begin{align}\label{eq:fcn_dab}
	\hspace{-0.2cm}\Delta_{AB}\!\!: \!\Tr \!\times \!\R^{nT L}&\times \R^{mTL}\to \R^{n^2T}\times \R^{nmT}\\
	\traj, \!\bDx\!, \!\bDu  \!&\longmapsto \! \Delta_{AB}(\traj, \!\bDx, \!\bDu) \!\coloneqq \!
	(\bDA, \!\bDB).\nonumber
\end{align}
Denote the set of well-conditioned batches in \eqref{eq:full_rank_batches} as
\begin{equation}\label{eq:FM}
\mathcal{F}_M \coloneqq \left\{ F \in \R^{(n+m)\times L}:
\kappa\left(FF^\top
\right)< M \right\}.
\end{equation}
In the following lemma, we characterize some properties of this set (the proof is provided in Appendix~\ref{proof:open_cone}). 
\vspace{0.2cm}
\begin{lemma}\label{lemma:open_cone}
$\mathcal{F}_M$ is an open cone in $\R^{(n+m)\times L}$.
\end{lemma}
\vspace{0.2cm}

\noindent Letting $\mathcal{F}_M^{T}\coloneqq \mathcal{F}_M\times \ldots \times  \mathcal{F}_M\subset \R^{nT L}\times \R^{mTL}$, the next lemma provides formal guarantees for the smoothness of $\Delta_{AB}$.
\begin{lemma}\label{lemma:error_on_dynamics} 
Let Assumption~\ref{hp:f_ell_regularity} 
hold. If $(\bDx, \bDu)\in \mathcal{F}_M^{T}$, then $\Delta_{AB}$ in~\eqref{eq:fcn_dab} is a $C^1$ function of $\traj$ and $\bDx, \bDu$. Furthermore, for each compact $\mathcal{H}\times \mathcal{K}\subset \Tr\times \mathcal{F}_M^{T}$, there exists $g(\mathcal{H}, \mathcal{K})>0$ such that, if $(\traj, \bDx, \bDu)\in\mathcal{H}\times\mathcal{K}$, then
\begin{equation}\label{eq:linear_bound_on_delta_ab}
\begin{split}
	\|\Delta_{AB}(\traj, \bDx, \bDu)\| \leq g(\mathcal{H}, \mathcal{K}) \|(\bDx, \bDu)\|.
\end{split}
\end{equation}
\end{lemma}
The proof is provided in Appendix~\ref{proof:lemma:error_on_dynamics}.
\begin{remark}\label{remark:openness}
Assumption \ref{hp:well_posed} ensures that it is possible to achieve the well-conditioning of gathered data in a neighbourhood of the optimal solution, for arbitrarily small dithers. Notice furthermore that, since $\Delta_{XU}$ is continuous and $\mathcal{F}_M^T$ is open by Lemma \ref{lemma:open_cone}, under Assumption \ref{hp:well_posed}, the preimage of $\mathcal{F}_M^T$ under $\Delta_{XU}$ is a nonempty open set.

\end{remark}

We now consider the approximated problem~\eqref{prob:pronto_approx} and the full knowledge problem~\eqref{eq:pronto_descent}.
Notice that, for a given trajectory $\traj$, these two problems differ only in the constraint represented by the LTV dynamics.
The descent direction error $\Delta_\zeta:=\hat{\zeta} - \zeta$ can be expressed as function of both the current trajectory $\traj$ and the estimation errors~\eqref{eq:bDAB}, so we define
\begin{equation}\label{eq:fcn_dzeta}
\begin{split}
\Delta_\zeta: &\Tr\times \R^{n^2T}\times \R^{nmT}\to \R^{s}\\
&\traj, \bDA, \bDB \longmapsto \Delta_\zeta(\traj, \bDA, \bDB)\coloneqq \hat{\zeta} - \zeta,
\end{split}
\end{equation}
where, for a given trajectory $\traj$, $\hat{\zeta}=(\hbdeltax, \hbdeltau)$ and $\zeta=(\bdeltax, \bdeltau)$  are the solutions of~\eqref{prob:pronto_approx} and~\eqref{eq:pronto_descent}, respectively.

\noindent The next lemma provides formal guarantees for this relation.
\begin{lemma}\label{lemma:solution_distances}
Let Assumption \ref{hp:f_ell_regularity} and \ref{hp:newton} hold. There exists a continuous and positive function $\delta_{AB}:\Tr\to \R_{>0}$, such that, if $\|(\bDA, \bDB)\|\leq \delta_{AB}(\traj)$, then $\Delta_\zeta$ in~\eqref{eq:fcn_dzeta} is a $C^1$ function of $\traj, \bDA, \bDB$ such that 
\begin{equation}
\Delta_\zeta(\traj, 0, 0) = 0.
\end{equation}
\end{lemma}
The proof is provided in Appendix~\ref{proof:lemma:solution_distances}.\\
In other words, Lemma~\ref{lemma:solution_distances} provides us with a trajectory-dependent bound $\delta_{AB}(\traj)$ on the identification error which, if respected, ensures that the descent error $\Delta_\zeta(\traj, \bDA, \bDB)$ is smooth in its arguments.

\subsection{Proof of Theorem~\ref{theo:main_result}}
\label{sec:proof}
We are now ready to prove the main result of this paper.
The proof goes through three main steps. \\
\textbf{Step I)} We show that, in a neighbourhood of the optimal solution $\traj^\star$ and for sufficiently small exploration dithers, it is possible to apply Lemma \ref{lemma:solution_distances}. \\
\textbf{Step II)} We leverage the smoothness properties of
functions $\Delta_{XU}, \Delta_{AB}, \Delta_\zeta$ in
\eqref{eq:fcn_dxu}, \eqref{eq:fcn_dab} and
\eqref{eq:fcn_dzeta} stated in Lemmas \ref{lemma:bounded_perturbations},~\ref{lemma:error_on_dynamics} and~\ref{lemma:solution_distances} respectively to obtain a linear bound on the descent direction error.\\
\textbf{Step III)}: We use this bound to guarantee practical
convergence of Algorithm~\ref{alg:DD-PRONTO} to $\traj^\star$.

\paragraph*{I) $\Delta_\zeta$ as smooth function of $\bdx, \bdu$}
Consider an isolated local minimum $\traj^\star$ of \eqref{prob:optcon_problem} and  
the function $\delta_{AB}(\cdot)$
given in Lemma~\ref{lemma:solution_distances}. 
Since $\delta_{AB}(\traj^\star)>0$, by continuity and strict positivity of $\delta_{AB}(\cdot)$ over all its domain, there exists $\sigma>0$ such that
\begin{equation}\label{eq:deltaabbound}
\delta_{AB}\coloneqq \min_{\traj \in \bm{\text{B}}_\sigma(\traj^\star)\cap \Tr}\delta_{AB}(\traj)>0.
\end{equation}
So, by Lemma \ref{lemma:solution_distances} it holds that for all $\traj \in \bm{\text{B}}_{\sigma}(\traj^\star)\cap \Tr$, if $\|(\bDA, \bDB)\|\leq \delta_{AB}$, then $\Delta_\zeta(\traj, \bDA, \bDB)$ is a $C^1$ function of its arguments.

\noindent
Pick any $\bm{\delta}_{xu}>0$ and $\sigma^\prime\in (0,\sigma]$, and consider triples $(\traj, \bdx, \bdu)$ in the set
\begin{equation}\label{eq:interesting_set}
\mathcal{H}\times \mathcal{D}\coloneqq \Big(\bm{\text{B}}_{\sigma^\prime}(\traj^\star)\cap \Tr \times \bm{\text{B}}_{\bm{\delta}_{xu}}\Big)\cap\Delta_{XU}^{-1}(\mathcal{F}_M^T),
\end{equation}
where $\Delta_{XU}^{-1}(\mathcal{F}_M^T)$ is the preimage of $\mathcal{F}_M^T$ under $\Delta_{XU}$ and $\sigma^\prime$ is chosen such that $\mathcal{H}\times \mathcal{D}$ is a compact set (using the same arguments of Remark \ref{remark:openness}, under Assumption \ref{hp:well_posed}, $\Delta_{XU}^{-1}(\mathcal{F}_M^T)$ is nonempty, open, and it is possible to find $\sigma^\prime$ and $\bm{\delta}_{xu}$ for which the resulting intersection in \eqref{eq:interesting_set} is compact).

\noindent
Under Assumptions~\ref{hp:f_ell_regularity} and~\ref{hp:pi_regularity}, we can apply Lemma~\ref{lemma:bounded_perturbations}, from which, since $\Delta_{XU}$ is $C^1$ in its arguments and $\Delta_{XU}(\traj, 0, 0)=0$ for all $\traj \in \Tr$, there exists a positive scalar $r=r(\sigma^\prime, \bm{\delta}_{xu})>0$ such that
\begin{equation}\label{eq:interesting_set_2}
\|(\bDx, \bDu)\| \leq r\|(\bdx, \bdu)\|
\end{equation}
for all $(\traj, \bdx, \bdu)$ satisfying \eqref{eq:interesting_set}. 
Next, since all triples in \eqref{eq:interesting_set} are mapped by $\Delta_{XU}$ into $(\bDx, \bDu) \in \mathcal{F}^T_M$ (by construction)
and since by \eqref{eq:interesting_set_2} it holds $(\bDx, \bDu) \in \bm{\text{B}}_{r\bm{\delta}_{xu}}$,
we apply Lemma \ref{lemma:error_on_dynamics} and obtain that there exists
$g=g(\sigma^\prime, \bm{\delta}_{xu})>0$ such that  
\begin{equation}\label{eq:interesting bound}
\begin{split}
	\|(\bDA, \bDB)\| \leq& g \|(\bDx, \bDu)\|\\
	\leq &gr \|(\bdx, \bdu)\|
\end{split}
\end{equation}
for all $(\traj, \bdx, \bdu)$ satisfying \eqref{eq:interesting_set}.
From \eqref{eq:interesting bound}, we obtain that for $\|(\bdx, \bdu)\| \leq \frac{\delta_{AB}}{gr}$, with $\delta_{AB}$ as in \eqref{eq:deltaabbound}, it follows from Lemma \ref{lemma:solution_distances} that $\Delta_\zeta(\traj, \Delta_{AB}(\traj, \Delta_{XU}(\traj, \bdx, \bdu)))$ is a $C^1$ function of its arguments.

\paragraph*{II) Bound for the maximum error on the descent direction}
Denote $\bm{\delta}_{xu}^\prime\coloneqq \min(\bm{\delta}_{xu}, \frac{\delta_{AB}}{gr})$.  Consider the composition map from the trajectory $\traj$ and exploration dithers $\bdx, \bdu$ to the descent error $\hat{\zeta}-\zeta$, namely
\begin{equation}\label{eq:compositionmap}
\hat{\zeta}-\zeta = \Delta_\zeta(\traj, \Delta_{AB}(\traj, \Delta_{XU}(\traj, \bdx, \bdu))),
\end{equation}
by applying Lemmas~\ref{lemma:bounded_perturbations},~\ref{lemma:error_on_dynamics} and~\ref{lemma:solution_distances}, we have that for triples $(\traj, \bdx, \bdu)$ in the compact set
\begin{equation}\label{eq:interesting_set_3}
\mathcal{H}\times \mathcal{D}^\prime\coloneqq\Big(\bm{\text{B}}_{\sigma^\prime}(\traj^\star)\cap \Tr \times \bm{\text{B}}_{\bm{\delta}_{xu}^\prime}\Big)\cap\Delta_{XU}^{-1}(\mathcal{F}_M^T),
\end{equation}
by continuous differentiability of $\Delta_{\zeta}$ and being $\Delta_\zeta(\traj, 0, 0)=0$ for all $\traj \in \Tr$, there exists a positive scalar $p=p(\sigma^\prime, \bm{\delta}_{xu}^\prime)>0$ such that
\begin{equation}\label{eq:delta_zeta_bounds}
\begin{split}
	\|\hat{\zeta}-\zeta\| &\leq p \|(\bDA, \bDB)\|\\
	&\leq pgr\|(\bdx, \bdu)\|.
\end{split}
\end{equation}

\paragraph*{III) Practical convergence based on descent-direction bounds} 

Notice that Algorithm~\ref{alg:DD-PRONTO} (namely, \algname/) is a perturbed version of Algorithm~\ref{alg:PRONTO} (\texttt{PRONTO}), which can be written as the dynamical system~\eqref{eq:perturbed_pronto_dynamics}. 
Specifically, at each algorithm iteration $\iter$, the descent error $\Delta \zeta^\iter$ in \eqref{eq:perturbed_pronto_dynamics} is given as a function of $\traj^\iter$ and $\iter$ by the composition~\eqref{eq:compositionmap}.

From Lemma \ref{lemma:practical_stability}, there exist a maximum descent error $\bar{\delta}_\zeta>0$, a class $\mathcal{KL}$ function $\phi(\cdot, \cdot)$, and a class $\mathcal{K}$ function $b(\cdot)$ such that, if $\|\Delta \zeta^\iter\| \leq\bar{\delta}_\zeta$ at each iteration, then the algorithm evolution satisfies \eqref{eq:KL_bound}. 
Define
\begin{equation}
q(\sigma^\prime)\coloneqq \sup\{q^\prime> 0: \phi(q^\prime, 0) < \sigma^\prime\}.
\end{equation}
By construction of $q$ and since $\mathcal{KL}$ functions are decreasing in the second argument, for any algorithm initialization $\traj^0\in \mathbb{B}_{q(\sigma^\prime)}(\traj^\star)\cap \Tr$, if $\|\Delta \zeta^\iter\| \leq\bar{\delta}_\zeta$ is satisfied at each iteration, the algorithm evolution is constrained to the ball $\mathbb{B}_\sigma(\traj^\star)$, namely, $\traj^\iter\in  \mathbb{B}_\sigma(\traj^\star)$ for all $\iter$.
Furthermore, thanks to the projection step in \eqref{eq:perturbed_pronto_dynamics}, it holds $\traj^\iter\in \Tr$ for all $\iter$.
Notice that, since $\bdx, \bdu$ in \eqref{eq:bdx_bdu_def} are stacks of exploration dithers, under Assumption \ref{hp:well_posed} it holds that
\begin{equation}\label{eq:delta_xu_bounds}
\|(\bdx^\iter, \bdu^\iter)\| \leq T (\delta_u + \delta_x)
\end{equation}
for all $\iter \in \N$, where $\delta_x$ and $\delta_u$ are the user-defined parameters of Algorithm \ref{alg:DD-PRONTO}. 
Furthermore, by assumption the dithers $(\bdx^\iter, \bdu^\iter)$ are chosen such that $\eqref{eq:full_rank_batches}$ holds at each iteration.
So, we are considering triples $(\traj, \bdx, \bdu)$ in the set
\begin{equation}
\Big(\bm{\text{B}}_{q}(\traj^\star)\cap \Tr \times \bm{\text{B}}_{T (\delta_u + \delta_x)}\Big)\cap\Delta_{XU}^{-1}(\mathcal{F}_M^T).
\end{equation}
Pick any $\delta_x, \delta_u>0$ such that $\delta_u + \delta_x \leq \bm{\delta}_{xu}^{\prime}/T$.
From \eqref{eq:delta_zeta_bounds} and \eqref{eq:delta_xu_bounds}, it holds
\begin{equation}\label{eq:final_bound}
\|\hat{\zeta}^\iter-\zeta^\iter\| \leq  \delta_\zeta(\delta_x, \delta_u)\coloneqq p grT(\delta_u + \delta_x)
\end{equation}
for all $k \in \N$, for all $\traj \in \bm{\text{B}}_{q}(\traj^\star)\cap \Tr$.
In this region, we can use the bound in \eqref{eq:final_bound} to ensure that $\|\Delta\zeta^\iter\| <\bar{\delta}_\zeta$.
Specifically, by picking $\delta_x, \delta_u>0$ such that
\begin{equation}
\delta_u + \delta_x \leq \min\left(\frac{\bm{\delta}_{xu}^\prime}{T} , \frac{\bar{\delta}_\zeta}{pgrT} \right),
\end{equation}
it holds from \eqref{eq:final_bound} that
\begin{equation}\label{eq:bound_luub}
\|\Delta\zeta^\iter\|\leq \delta_\zeta(\delta_x, \delta_u) \leq  \bar{\delta}_\zeta
\end{equation}
for all $\iter\in \N$.
The bound in~\eqref{eq:bound_luub} then ensures that Algorithm~\ref{alg:DD-PRONTO} is Locally Uniformly Ultimately Bounded by Lemma \ref{lemma:practical_stability}.
Finally, we prove also that the ultimate bound in Theorem \ref{theo:main_result} is given by a strictly increasing function of $\delta_x, \delta_u$.
Since the ultimate bound $b(\delta_\zeta)$ from Lemma \ref{lemma:practical_stability} is a class $\K$ function of its argument $\delta_\zeta$, and since $\delta_\zeta(\delta_x, \delta_u)$ in \eqref{eq:final_bound} is linear in its arguments, the result follows.

\section{Numerical Example}\label{sec:simulations}

In this section, we demostrate the capabilities of \algname/ by
solving an optimal control problem for a nonlinear underactuated
robot, the so-called pendubot~\cite{siciliano2009force, zhang2002hybrid}, with unknown dynamics.
First, we present the setup and the nonlinear optimal control problem, then we 
compare our \algname/ (Algorithm~\ref{alg:DD-PRONTO}) with its
inspiring, model-based method (Algorithm~\ref{alg:PRONTO}).

\begin{figure}[h]
\centering
\hspace{-0.3cm}
\includegraphics[scale = 0.7]{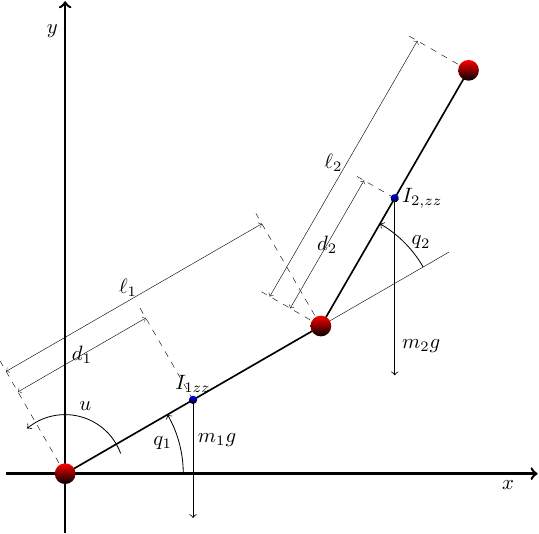}
\caption{The pendubot.}
\label{fig:pendubot}
\end{figure}

\noindent The pendubot consists of two links and one actuator on the first joint (see Figure~\ref{fig:pendubot}).
Its dynamics read as
\begin{align}
\label{eq:robot}
M(q)
\begin{bmatrix}
	\ddot{q}_1 \\ \ddot{q}_2
\end{bmatrix}
+ (C(q, \dot{q}) + F)
\begin{bmatrix}
	\dot{q}_1 \\ \dot{q}_2
\end{bmatrix} + G(q) = \begin{bmatrix}
	u \\ 0
\end{bmatrix},
\end{align}
where $q = (q_1, q_2) \in [0, 2\pi]^2$ stacks the two joint angles $q_1, q_2 \in \R$, 
$u \in \R$ is the input torque on the first joint,
$M(q) \in \R^{2\times 2}$ is the inertia matrix, $F \in \R^{2\times 2}$ accounts for friction, $C(q, \dot{q}) \in \R^2$ includes the Coriolis and centrifugal forces and $G(q) \in \R^2$ is the gravitational term. 
The matrices in~\eqref{eq:robot} are defined as
\begin{equation}
\begin{split}
	M(q) &:= 
	\begin{bmatrix}
		a_1 + a_2 + 2a_3\cos(q_2) & a_2 + a_3\cos(q_2)\\
		a_2 + a_3\cos(q_2) & a_2
	\end{bmatrix}\\
	C(q, \dot{q}) &  :=
	\begin{bmatrix}
		-a_3 \sin(q_2)\dot{q}_2  & -a_3 \sin(q_2)(\dot{q}_1 + \dot{q}_2)\\
		a_3 \sin(q_2)\dot{q}_1& 0
	\end{bmatrix}\\
	G(q) &  :=
	\begin{bmatrix}
		a_4\cos(q_1) + a_5\cos(q_1+q_2)\\
		a_5 \cos(q_1+q_2)
	\end{bmatrix}\\
	F  & \coloneqq \diag(f_1, f_2)
\end{split}
\end{equation}
where
\begin{equation}
\begin{split}
	a_1 &=  I_{1, zz} + m_1d_1^2  + m_2\ell_1^2\\
	a_2 &= I_{2, zz} + m_2 d_2^2, \hspace{1cm} a_3 = m_2 \ell_1 d_2\\
	a_4 &= g(m_1 d_1 + m_2 \ell_1), \hspace{1cm} 	a_5 = gm_2 d_2.
\end{split}
\end{equation}
The physical parameters used to simulate the system are summarized in Table \ref{table:params}.
\begin{table}
\caption{Physical parameters of the pendubot.}
\begin{center}
	\begin{tabular}{c | c | c | c}\label{table:params}
		Parameter & Value & Parameter & Value \\
		\hline
		$m_1$[Kg] & $1$& $m_2$[Kg] & $1$ \\
		$\ell_1$[m] & $1$ & $\ell_2$[m] & $1$\\
		$d_1$[m] & $0.5$& $d_2$[m] & $0.5$ \\
		$I_{1, zz}$[Kg$m^2$] & $0.33$  & $I_{2, zz}$[Kg$m^2$] & $0.33$ \\
		$f_1$[Ns/rad] & $0.1$ & $f_2$[Ns/rad] & $0.1$ \\
	\end{tabular}
\end{center}
\end{table}
From~\eqref{eq:robot}, we obtain a state space model with state variable $x = (q_1, q_2, \dot{q}_1, \dot{q}_2)$.
The dynamics is then discretized via forward Euler integration of step $dt=0.01$s over a simulation time of $T=10$s. 
The cost function is a quadratic cost function designed to follow a step reference $(\bx^*, \bu^*)$ describing a swing up:
\begin{equation}
\begin{split}
	\ell(\bx, \bu) =& \sum_{t=0}^{T-1} (x_t\!-\!x^*_t)^\top \!Q(x_t\!-\!x^*_t) \!+(u_t\!-\!u^*_t)^\top \!R (u_t\!-\!u^*_t) \\
	&+(x_T\!-\!x^*_T)^\top \!Q_T(x_T\!-\!x^*_T),
\end{split}
\end{equation} 
with $Q = \diag(10^3,10^3,10^2,10^2)$, $R=50$ and $Q_T$ found by solving the discrete-time algebraic Riccati equation on the final vertical equilibrium.
The reference state curve is a step from an initial (stable) equilibrium $x_0 = (-\frac{\pi}{2}, 0, 0, 0)$ to the final (unstable) equilibrium $x_T = (\frac{\pi}{2}, 0, 0, 0)$. The reference input curve compensates for the gravity term at the two equilibrium position. 
The initial trajectory $\traj^0$ for the algorithm is chosen as the standstill robot in the starting position.

\subsection{Pendubot swing-up numerical experiment}

\noindent
The data-driven controller $\pi$ is approximated about a given trajectory by
designing at each iteration $\iter$ a finite-horizon LQR problem over
an inexact linearization of the current nominal trajectory
$\traj^\iter$. That is, we rely on the estimated $\hat{A}_t, \hat{B}_t$, which are used
both to find the descent direction and to obtain an estimated
controller.

\begin{figure}[h]
	\hspace{-0.3cm}
	\centering
	\includegraphics[scale = 0.8]{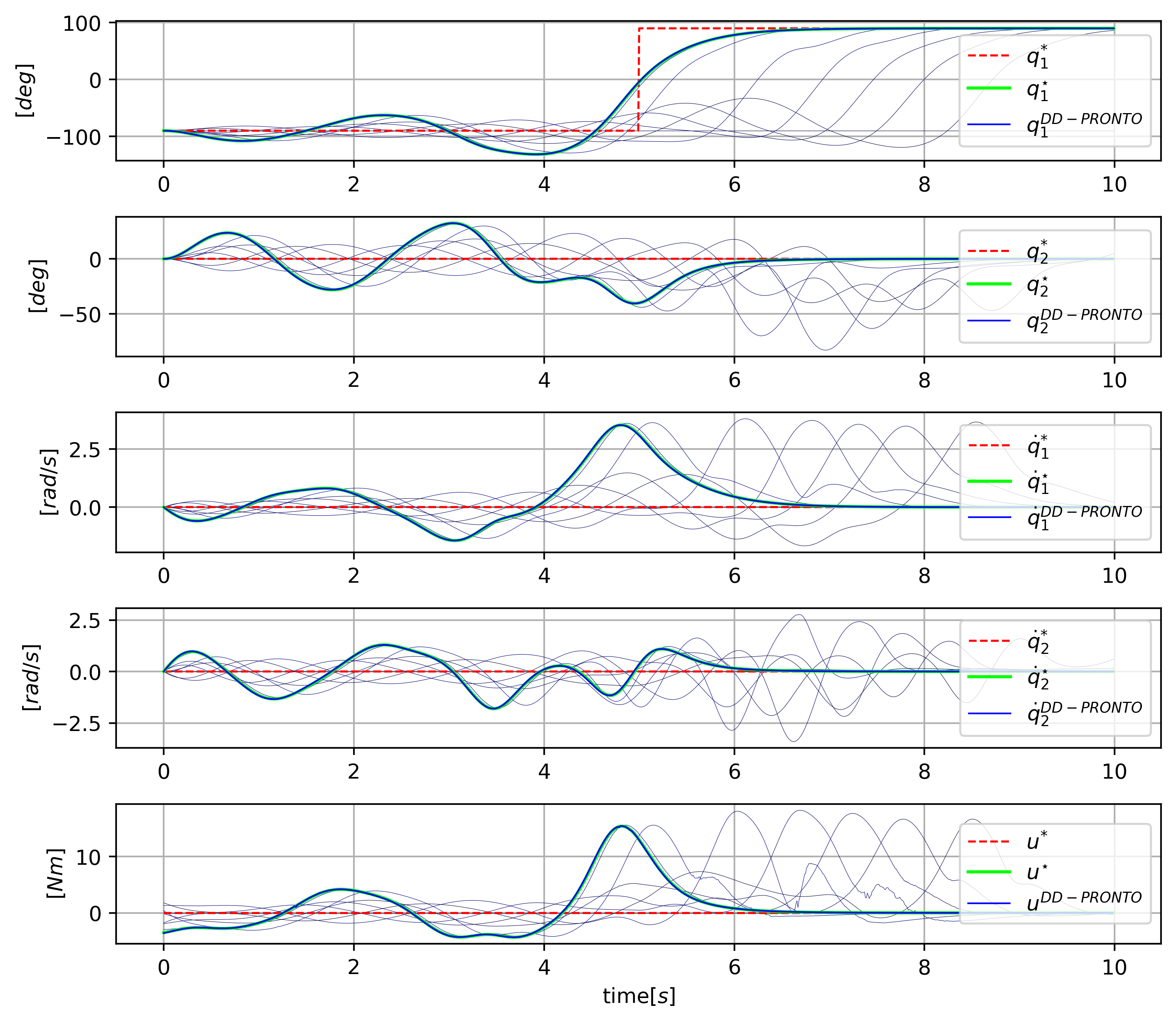}
	\caption{In blue the reference curves for the states and the input. In red, the result of \algname/. In green, the optimal trajectory.}
	\label{fig:trajectory}
\end{figure}

\begin{figure}[h]
	\centering
	\includegraphics[scale = 0.8]{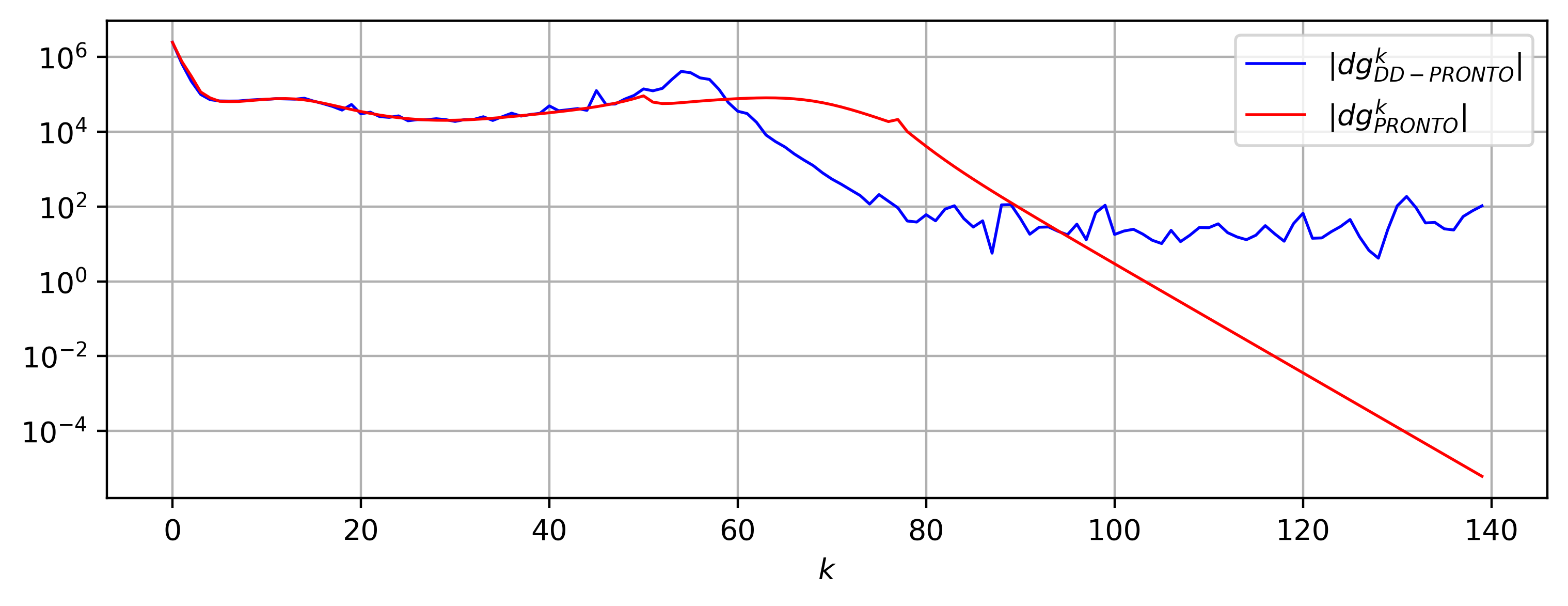}
	\caption{Evolution of \texttt{PRONTO} (in red) and \algname/ (in blue).}
	\label{fig:descent}
\end{figure}

\noindent
In order to collect the perturbations $\hat{\traj}^{i, \iter}$ of the current trajectory $\traj^\iter$ at each iteration $\iter$, we add the exploration dither as in~\eqref{eq:closed_loop_experiment}, with $d^{i, \iter}_{u, t}\sim \mathcal{U}(0, \delta_u)$ and $d^{i, \iter}_{x, t}\sim \mathcal{U}(-\delta_x,\delta_x)$, with $\delta_u = 0.1, \delta_x=0.01$. At each iteration, we collect $L=6$ perturbed trajectories. 
The stepsize is $\gamma=1$.
\noindent 
In Figure~\ref{fig:trajectory}, we plot the iterations of \algname/ for all the states and the input.
In Figure~\ref{fig:descent}, we plot the absolute value of the cost differential 
\begin{equation}
\text{d}g(\traj^\iter)\coloneqq \frac{\text{d}g}{\text{d}\traj}\Big|_{\traj^\iter}\zeta^\iter = \sum_{t=0}^{T-1}\left(q_t^{\iter \top}\Delta x_t^\iter + r_t^{\iter \top} \Delta u_t^\iter\right) + q_T^{\iter \top} \Delta x_T^\iter
\end{equation}
along the evolution of Algorithm \ref{alg:PRONTO} (in red) and
Algorithm \ref{alg:DD-PRONTO} (in blue). Ultimate boundedness of
\algname/ is highlighted by the fact that it does not reach
the optimal solution and the cost differential $\text{d}g(\traj^\iter)$ never
reaches zero.


\subsection{Dither amplitude and suboptimality}

In order to better see the convergence properties of \algname/ depending on the parameters $\delta_x, \delta_u$, we test $15$ times the algorithm convergence, solving the same optimal control problem.
To do this, we halve at each algorithm instance $j$ the bound on the exploration noise, namely, $\delta_x^{j+1}=\delta_x^j/2$ and $\delta_u^{j+1}=\delta_u^j/2$. We choose at the first algorithm run $\delta_{x}^0=0.01$ and $\delta_u^0=0.1$, and we pick $d^{i, \iter}_{x, t}\sim \mathcal{U}(-\delta^j_x,\delta^j_x)$ and $d^{i, \iter}_{u, t}\sim \mathcal{U}(0, \delta^j_u)$, for all $t$.
The control law $\pi$ is chosen in a data-driven way as detailed in the previous section.

\begin{figure}[h]
\centering
\includegraphics[scale = 0.8]{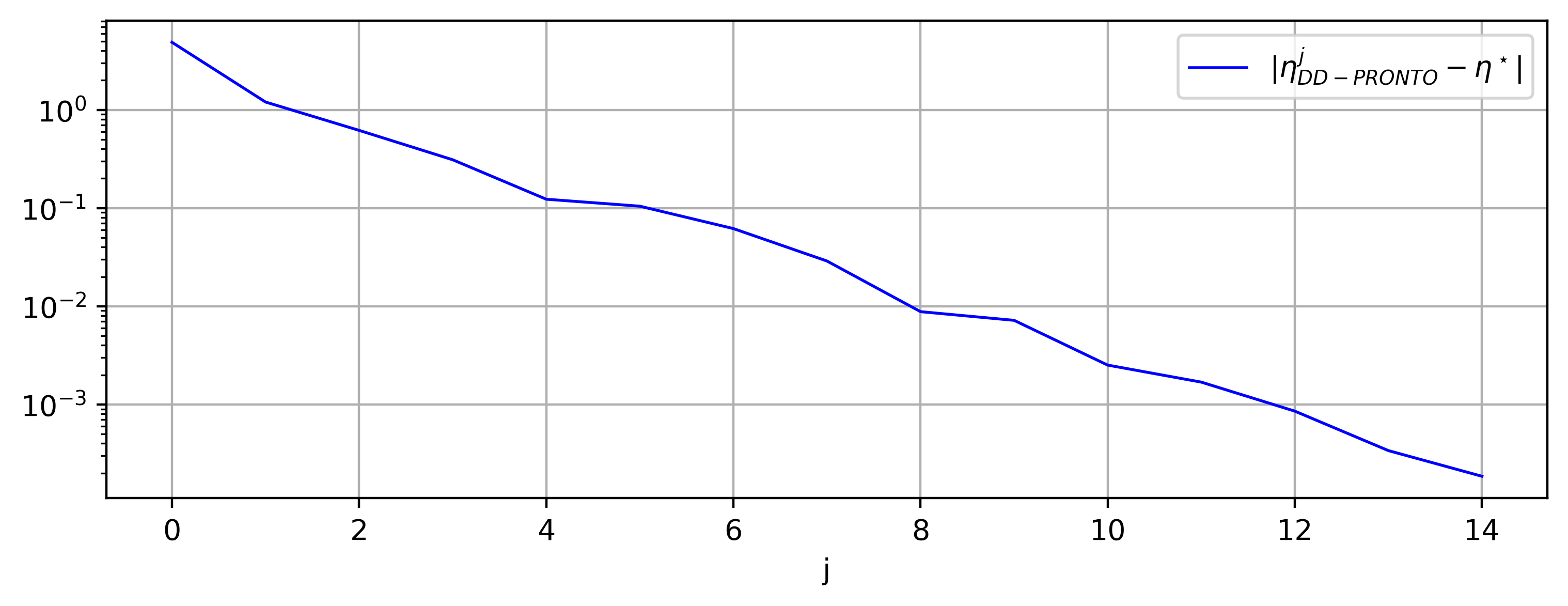}
\caption{Distance from the optimum depending on the dither amplitude.}
\label{fig:ultimate_distance_wrong}
\end{figure}

\noindent
The results, showed in Figure
\ref{fig:ultimate_distance_wrong}, demonstrate the strictly decreasing
suboptimality of \algname/ as the dither amplitude decreases.  

\section{Conclusions}\label{sec:conclusions}

In this paper we proposed a novel data-driven optimal control algorithm called \algname/. This algorithm extends the applicability of \texttt{PRONTO} by removing the knowledge of the system model, relying on the ability to explore system trajectories. 
The main advantage of this algorithm is the possibility to overcome the suboptimality of model-based solution by estimating the linearizations about a trajectory instead of computing them from the model. Theoretical guarantees on the convergence of the algorithm have been given, together with insight on how to tune the design parameters of the algorithm.

\section{Appendix}\label{sec:appendix}

\subsection{Proof of Lemma \ref{lemma:hessian_approximation_convergence}}
\label{proof:lemma:hessian_approximation_convergence}
We divide the proof in three steps. First, we rewrite the locally exponentially stable optimization algorithm \cite[Algorithm $4.1$]{diehl2011numerical} with Hessian approximation in a form which highlights the analogies with the dynamical system \eqref{eq:approx_pronto_dynamics}.
Next, we show that introducing (in \cite[Algorithm $4.1$]{diehl2011numerical}) a projection step onto the trajectory manifold $\Tr$ preserves local exponential stability. Third, we show that the obtained projected SQP dynamics is indeed dynamics \eqref{eq:approx_pronto_dynamics}.

\paragraph*{I) Approximate Newton method dynamics}

Consider problem~\eqref{prob:h_formulation}. The (approximate) Newton method for constrained optimization applied to~\eqref{prob:h_formulation} updates the tentative solution $\curve^\iter$ by applying, for all iterations $\iter$, the iterative rule, cf.~\cite[Algorithm $4.1$]{diehl2011numerical},
\begin{equation}
\label{eq:newtonupdate}
\begin{bmatrix}
	\curve^{\iter+1}\\
	\lambda^{\iter+1}
\end{bmatrix}
=
\begin{bmatrix}
	\curve^{\iter}\\
	0
\end{bmatrix}
-\gamma
\underbrace{\begin{bmatrix}
		\dd^2 \ell(\curve^\iter) &\dd h(\curve^\iter)\\
		\dd h(\curve^\iter) ^\top & 0
	\end{bmatrix}^{-1}}_{Z(\curve^\iter)}
\begin{bmatrix}
	\dd \ell(\curve^\iter)\\
	h(\curve^\iter),
\end{bmatrix}
\end{equation}
where $\lambda^\iter \in \R^{s_x}$ are the lagrangian multipliers associated to the equality constraint. This means the update law for $\curve^{\iter}$ is given by
\begin{equation}\label{eq:xi_update}
\begin{split}
	\zeta^\iter &= \underbrace{Z_{11}(\curve^\iter) \dd \ell(\traj^\iter)}_{\coloneqq \zeta_1^\iter} + \underbrace{Z_{12}(\curve^\iter)h(\curve^\iter)}_{\coloneqq \zeta_2^\iter}\\
	\curve^{\iter+1} &= \curve^\iter + \gamma\zeta^\iter.
\end{split}
\end{equation}
where $Z_{ij}$ denotes the $ij$-th block of matrix $Z$ in~\eqref{eq:newtonupdate}.
It can be proved \cite[Thm. 4.2]{diehl2011numerical} that, under Assumptions, \ref{hp:newton} and \ref{hp:f_ell_regularity}, a local solution $\traj^\star$ to \eqref{prob:h_formulation} is LES for dynamics \eqref{eq:xi_update}. Notice that in general the tentative solution $\curve^\iter$ is not a trajectory of the controlled system.
It can be shown (by writing the KKT conditions of \eqref{eq:zeta_1}) that the term $\zeta_1^\iter$ in~\eqref{eq:xi_update} is the result of the following optimization problem
\begin{equation}\label{eq:zeta_1}
\begin{split}
	\zeta_1^\iter = \argmin_{\zeta} \;\;\;& \zeta^\top \dd^2 \ell(\curve^\iter) \zeta + \dd \ell(\curve^\iter)^\top\zeta \\
	\text{s.t. }\;\;& \dd h(\curve^\iter)^\top \zeta = 0,
\end{split}
\end{equation}
namely, it satisfies by construction $\zeta_1^\iter \in T_{\xi^\iter} \Tr$ at all iterations. The second term, $\zeta_2^\iter$, takes into account the constraint violation, and it is zero when $\curve^\iter \in \Tr$.

\paragraph*{II) Introducing the projection}
We show now that, introducing a projection, the update law \eqref{eq:xi_update} does not lose its convergence properties. Consider the update law
\begin{equation}\label{eq:nu_update_0}
\begin{split}
	\zeta^\iter &= Z_{11}(\curve^\iter) \dd \ell(\curve^\iter) + Z_{12}(\curve^\iter)h(\curve^\iter)\\
	\curve^{\iter+1} &= \pro(\curve^\iter + \gamma\zeta^\iter).
\end{split}
\end{equation}
Thanks to the projection $\pro$ in the update, we know it holds $\curve^\iter\in \Tr$ for all $\iter \in \N$. For this reason, we re-write \eqref{eq:nu_update_0} as
\begin{equation}\label{eq:nu_update}
\begin{split}
	\zeta^\iter &= Z_{11}(\traj^\iter) \dd \ell(\traj^\iter) + Z_{12}(\traj^\iter)h(\traj^\iter)\\
	\traj^{\iter+1} &= \pro(\traj^\iter + \gamma\zeta^\iter),
\end{split}
\end{equation}
where we only highlithted by using $\traj^\iter$ the fact that iteration \eqref{eq:nu_update}, unlike \eqref{eq:xi_update}, produces only system trajectories.
Under Assuption \ref{hp:pi_regularity}, we can expand the update in Taylor series and we obtain
\begin{equation}\label{eq:nu_update_2}
\begin{split}
	\zeta^\iter &= Z_{11}(\traj^\iter) \dd \ell(\traj^\iter) + Z_{12}(\traj^\iter)h(\traj^\iter)\\
	\traj^{\iter+1} &= \pro(\traj^\iter) + \gamma\dd\pro(\traj^\iter)^\top \zeta^\iter + o(\zeta^\iter)\\
	&= \traj^\iter + \gamma\dd\pro(\traj^\iter)^\top \zeta^\iter+ o(\zeta^\iter),\\
	&= \traj^\iter + \gamma\dd\pro(\traj^\iter)^\top (\zeta_1^\iter + \zeta_2^\iter)+ o(\zeta^\iter),
\end{split}
\end{equation}
where we exploited the fact that $\pro(\traj)=\traj$ for all $\traj \in \Tr$.
It can be shown (see \cite{hauser2002projection} for more insight) that, if $\traj\in \Tr$ and $\zeta \in T_{\traj}\Tr$, it holds
\begin{equation}
\nabla \pro(\traj)^\top \zeta = \zeta,
\end{equation}
namely, $\nabla \pro^\top: \R^s \to T_{\traj}\Tr$ is itself a projection into the space tangent to the trajectory manifold at $\traj$. Since $\zeta_1^\iter\in T_{\traj^\iter}\Tr$ by construction (being the solution of \eqref{eq:zeta_1}), and since $\zeta_2^\iter = Z_{12}(\traj^\iter)h(\traj^\iter)=0$ (as $\traj\in \Tr\iff h(\traj)=0$), we re-write \eqref{eq:nu_update_2} as
\begin{equation}\label{eq:nu_update_3}
\begin{split}
	\zeta^\iter &= Z_{11}(\traj^\iter) \dd \ell(\traj^\iter) + Z_{12}(\traj^\iter)h(\traj^\iter)\\
	\traj^{\iter+1} &= \traj^\iter + \gamma\zeta^\iter + o(\zeta^\iter),
\end{split}
\end{equation}
Notice that dynamics \eqref{eq:xi_update} and \eqref{eq:nu_update_3} differ only in the little-o term.
By definition of $o(\zeta)$, for all $\epsilon>0$ we can find $\delta_\epsilon>0$ such that
\begin{equation}
\|\zeta\|\leq \delta_\epsilon \implies \|o(\zeta) \|\leq \epsilon \|\zeta \|.
\end{equation}
Furthermore, holding the second order sufficient condition of optimality for $\traj^\star$ and being $f, \ell$ smooth, it can be shown that for any ball $\mathbb{B}_r(\traj^\star)$, there exists $k(r)>0$ such that $\|\zeta\| \leq k(r)\|\traj-\traj^\star\|$ for all $\traj \in \mathbb{B}_r(\traj^\star)$. Overall, we have that for any $\epsilon>0$ there exists $\delta_\epsilon>0$ such that
\begin{equation}\label{eq:bound_for_little_o}
\|\traj-\traj^\star \| \leq \delta_\epsilon \implies \|o(\zeta(\traj))\| \leq \epsilon\|\traj-\traj^\star\|.
\end{equation}
In \cite[Thm. 4.2]{diehl2011numerical}, it is shown that by choosing the Lyapunov function $V(\xi)=\|\xi-\traj^\star\|$, the optimal solution $\traj^\star$ is Locally Exponentially Stable for dynamics \eqref{eq:xi_update}, and for some $s<1, r>0$ it holds
\begin{equation}
\begin{split}
	\|\curve^{\iter+1}-\traj^\star \| & = 	\|\curve^{\iter} + \gamma \zeta^\iter-\traj^\star \| \\
	&\leq s \| \curve^\iter - \traj^\star\|
\end{split}
\end{equation}
for all $\curve \in \mathbb{B}_r(\traj^\star)$. 
By using the same Lyapunov function for dynamics \eqref{eq:nu_update_3}, we obtain that it must hold
\begin{equation}
\begin{split}
	\|\traj^{\iter+1} - \traj^\star\| =& \|\traj^\iter + \gamma\zeta^\iter +o(\zeta) - \traj^\star \|\\
	\leq& s\|\traj^\iter -\traj^\star\| + \|o(\zeta)\|.
\end{split}
\end{equation}
Recalling \eqref{eq:bound_for_little_o}, picking $\epsilon: s+\epsilon<1$, we have that there exists $\delta_\epsilon>0$ such that
\begin{equation}\label{eq:lyap_decrease}
\begin{split}
	\|\traj^{\iter+1} - \traj^\star\| \leq  (s+\epsilon)\|\traj^\iter -\traj^\star\| 
\end{split}
\end{equation}
for all $\traj^\iter \in \mathbb{B}_{\delta_\epsilon}(\traj^\star)\cap \mathbb{B}_r(\traj^\star)$, which proves LES of $\traj^\star$ under dynamics \eqref{eq:nu_update_3}. 

\paragraph*{III) The obtained update rule is the same as in \eqref{eq:approx_pronto_dynamics}}
At last, notice that since $\traj^\iter \in \Tr$ for all $\iter$, we have $h(\traj^\iter)=0$ and thus the update \eqref{eq:nu_update_3} can be rewritten as
\begin{equation}
\begin{split}
	\zeta^\iter &= Z_{11}(\traj^\iter) \dd \ell(\traj^\iter)\\
	\traj^{\iter+1} &= \pro(\traj^\iter + \gamma\zeta^\iter),
\end{split}
\end{equation}
where $Z_{11}(\traj^\iter)$ is in \eqref{eq:newtonupdate}, and this dynamics is the same implemented by Algorithm~\ref{alg:DD-PRONTO} in \eqref{eq:approx_pronto_dynamics}. 
We can thus conclude Local Exponential Stability of $\traj^\star$ under dynamics \eqref{eq:approx_pronto_dynamics}\cite[Thm. 13.11]{haddad2008nonlinear}. 

\subsection{Proof of Lemma~\ref{lemma:practical_stability}}
\label{proof:lemma:practical_stability}

The proof is obtained by applying~\cite[Theorem 2.7]{cruz1999stability}.
First, notice that by taking the squares of \eqref{eq:lyap_decrease}, it is possible to use $V(\traj)\coloneqq \|\traj-\traj^\star\|^2$ as $C^1$ Lyapunov function for the unperturbed dynamics \eqref{eq:approx_pronto_dynamics}. This means that Assumption B1 of \cite[Theorem 2.7]{cruz1999stability} is satisfied since the equilibrium $\traj^\star$ of unperturbed system is locally asymptotically stable with $C^1$ Lyapunov function $V(\traj)=\|\traj-\traj^\star\|^2$.
Next, we rewrite the perturbed system \eqref{eq:perturbed_pronto_dynamics} as
\begin{align}\label{eq:ddpronto_dyn}
	\traj^{\iter+1} &= \pro( \traj^\iter + \gamma \zeta^\iter + \gamma \Delta \zeta^\iter) \\
	&= \underbrace{\pro(\traj^\iter \!+ \!\gamma \zeta^\iter)}_{\text{unperturbed}} + \underbrace{\gamma \dd \pro^\top_{\traj^\iter + \gamma \zeta^\iter} \Delta\zeta^\iter \!+ o\left((\Delta \zeta^\iter)^2\right)}_{\text{perturbation } d(\traj^\iter, \Delta\zeta^\iter)},\nonumber
\end{align}
for $\Delta \zeta \rightarrow 0$, to explicit the disturbance as an additive term.
Finally, to satisfy assumption B2 of \cite[Theorem 2.7]{cruz1999stability}, we need a bound for the perturbation term. 
Being $d(\traj, \Delta \zeta)$ differentiable in its arguments and such that $d(\traj, 0)=0$ for all $\traj \in \R^s$, for every ball $\mathbb{B}_r\subset \R^{s\times s}$ there exists $\bar{d}(r)>0$ such that
\begin{equation}
	\|d(\traj^\iter, \Delta \zeta^\iter)\| \leq \bar{d}(r) \| \Delta \zeta^\iter\|\;\;\; \forall \Delta \zeta^\iter\in \mathbb{B}_r, \traj^\iter \in \Tr.
\end{equation}
Denote now as $B\subset\R^s$ the basin of attraction of the unperturbed dynamics \eqref{eq:approx_pronto_dynamics}. 
Then, by \cite[Theorem 2.7]{cruz1999stability}, if $\|\Delta \zeta^\iter\|\leq \delta_\zeta$ for all $\iter \in \N$,
there exists $\bar{\delta}_\zeta>0$ such that, if $\traj^0 \in B$ and $\|\Delta \zeta^\iter \|\leq \delta_\zeta < \bar{\delta}_{\zeta}$ for all $\iter$, then there exists $N\in \N$, class $\KL$ function $\phi$ and class $\K$ function $b$ such that 
\begin{equation}
\begin{split}
	\| \Delta \traj^\iter\|&\leq\phi(\|x_0 \|, k) \hspace{1cm} \forall k<N\\
	\|\Delta \traj^\iter\|&\leq b(\delta_\zeta) \hspace{1.8cm}  \forall k\geq N.
\end{split}
\end{equation}

\subsection{Proof of Lemma~\ref{lemma:bounded_perturbations}}
\label{proof:lemma:bounded_perturbations}
Notice that each perturbation $\hat{\traj}^i=(\hat{\bx}^i,\hat{\bu}^i)$ is obtained via the closed-loop \eqref{eq:closed_loop_experiment}, which can be seen as a repeated composition of the functions $\pi$ (controller), $f$ (dynamics) and the dither injection.
We thus write $\hat{\traj}^i=\hat{\traj}^i(\traj, \bdx, \bdu)$, where the dependence on $\traj=(\bx, \bu)$ takes into account the fact that $\pi$ is tracking the current trajectory $\traj$.
By composition of $C^1$ functions, under Assumptions \ref{hp:f_ell_regularity}, \ref{hp:pi_regularity}, we have then that for all iterations $i=1, \ldots, L$, $\hat{\traj}^i(\traj, \bdx, \bdu)$ is a $C^1$ function of $\traj, \bdx, \bdu$. Next, notice that the data batches built as in \eqref{eq:batches} are stacks of differences between components of $\hat{\traj}^i$ and components of $\traj$, so the function $\Delta_{XU}$ can be seen as a smooth composition between functions $\hat{\traj}^i(\traj, \bdx, \bdu)$, with $i=1, \ldots, L$, and $\traj$, which means $\Delta_{XU}(\traj, \bdx, \bdu)$ is $C^1$. 
To conclude the proof, if all dithers $\bdu, \bdx$ are zero, then by Assumption \ref{hp:pi_regularity} we have that all perturbed trajectories $\hat{\traj}^i$ coincide with the current one $\traj$, namely, $\hat{\traj}^i(\traj, 0, 0)=\traj$ for all $i=1, \ldots, L$. In turn, this means that the difference $\hat{\traj}^i(\traj, 0, 0)-\traj$ used to build $\Delta_{XU}$ is zero, and so $\Delta_{XU}(\traj, 0, 0)=0$.

\subsection{Proof of Lemma~\ref{lemma:open_cone}}\label{proof:open_cone}
At first, we show $\mathcal{F}_M$ is open. 
Notice that any $F\in \mathcal{F}_M$ must be of full row rank, otherwise $\kappa(FF^\top)=\infty$ and $F\notin \mathcal{F}_M$. For any $F$, the map $FF^\top$ is continuous in $F$, and since $FF^\top$ is Hermitian with positive eigenvalues for all full row rank $F$, its condition number is given by
\begin{equation}
\begin{split}
	\kappa(FF^\top)&\coloneqq |\lambda_{\text{max}}(FF^\top)/\lambda_{\text{min}}(FF^\top)|\\
	&=\lambda_{\text{max}}(FF^\top)/\lambda_{\text{min}}(FF^\top),
\end{split}
\end{equation} 
where $\lambda_{\text{max}}(\cdot)$ and $\lambda_{{\text{min}}}(\cdot)$ are the maximum and minimum eigenvalues of a matrix,
which is a continuous function of $FF^\top$ for any full row rank $F\in \R^{(n+m)\times L}$. Thus, the map $F\mapsto \kappa(FF^\top)$ is continuous by composition of continuous maps for any $F\in \mathcal{F}_M$. 
Pick any $F\in \mathcal{F}_M$, and define $\epsilon_F = (M-\kappa(FF^\top))/2$. By continuity, we can always find $\delta_F>0$ such that
\begin{equation}
|F^\prime - F|< \delta_F \implies |\kappa(F^\prime F^{\prime \top})-\kappa(FF^\top)|< \epsilon_F,
\end{equation}
namely, $\kappa(F^\prime F^{\prime \top})< M$, and thus $F^\prime \in \mathcal{F}_M$. In other words, we can find an open ball about any point $F\in\mathcal{F}_M$ (whose radius depends on $F$) which is contained in the set $\mathcal{F}_M$.		

Next, we show $\mathcal{F}_M$ is a cone. 
This follows from the fact that if $F\in \mathcal{F}_M$, then also $\lambda F\in \mathcal{F}_M$, since
\begin{equation}
\begin{split}
	\kappa(\lambda^2 FF^\top )&=|\lambda_{\text{max}}(\lambda^2 FF^\top)/\lambda_{\text{min}}(\lambda^2 FF^\top)|\\
	&= \frac{\lambda^2}{\lambda^2}\kappa(FF^\top)\\
	&=\kappa(FF^\top)<M.
\end{split}
\end{equation}

\subsection{Proof of Lemma~\ref{lemma:error_on_dynamics}}
\label{proof:lemma:error_on_dynamics}

The proof goes through three main steps. 
First, we find a closed-form expression for the errors \eqref{eq:deltaA_deltaB} on the Jacobians at each $t$.
Second, we show that this expression is a differentiable function of the entries of $\Delta X_t, \Delta U_t$.
Finally, we define a local linear bound on the norm of $\Delta_{AB}$.

\paragraph*{I) Closed-form expression for $\Delta A_t$ and $\Delta B_t$.}

Define 
\begin{equation}
\begin{split}
	\Delta x^i_t &= \hat{x}^i_t-x_t,\;\;\;\Delta u^i_t = \hat{u}^i_t-u_t,\\
	\Delta x_t^{+,i} &= f(\hat{x}^i_t, \hat{u}^i_t) - f(x_t, u_t)
\end{split}
\end{equation}
where the apex $i$ denotes the $i$-th perturbation of the current trajectory $\traj=(\bx, \bu)$. 
By expanding in Taylor series the dynamics $f$ about $x_t, u_t$, we obtain:
\begin{equation}\label{eq:first_approx}
\Delta x^{+,i}_t=A_t\Delta x^i_t+ B_t\Delta u^i_t + o_\traj(\Delta x^i_t, \Delta u^i_t),
\end{equation}
for $\Delta x_t^i\rightarrow 0, \Delta u_t^i\rightarrow 0$. The pedex $\traj$ in $o_\traj(\cdot)$ highlights the fact that the little-o term depends on the trajectory $\traj$.
We then build $\Delta X_t, 	\Delta U_t, \Delta X^{+}_t$ as in \eqref{eq:batches}.
It holds:
\begin{equation}
\Delta X^{+}_t= \begin{bmatrix}
	A_t & B_t
\end{bmatrix}
\begin{bmatrix}
	\Delta X_t\\
	\Delta U_t
\end{bmatrix}
+ o_\traj\left(\begin{bmatrix}
	\Delta X_t\\
	\Delta U_t
\end{bmatrix}\right),
\end{equation}
for $\Delta x_t^i\rightarrow 0, \Delta u_t^i\rightarrow 0$ for all $i=1, \ldots, L$, from which, under Assumption \ref{hp:well_posed}, we obtain:
\begin{equation}\label{eq:taylor_expansion}
\begin{bmatrix}
	A_t & B_t
\end{bmatrix}=
\left(\Delta X_t^{+} - o_\traj\left(\begin{bmatrix}
	\Delta X_t\\
	\Delta U_t
\end{bmatrix}\right)\right)
\begin{bmatrix}
	\Delta X_t\\
	\Delta U_t
\end{bmatrix}^{\dagger}.
\end{equation}
Recall that we estimate the Jacobians with
\begin{equation}\label{eq:estimation}
\begin{bmatrix}
	\hat{A}_t & \hat{B}_t
\end{bmatrix}=
\Delta X_t^{+}
\begin{bmatrix}
	\Delta X_t\\
	\Delta U_t
\end{bmatrix}^{\dagger}.
\end{equation}
We can then subtract \eqref{eq:taylor_expansion} from \eqref{eq:estimation} to obtain 
\begin{align}\label{eq:expression_for_delta_AB}
\begin{bmatrix}
	\Delta A_t & \Delta B_t
\end{bmatrix}	
& = o_\traj\left(\begin{bmatrix}
	\Delta X_t\\
	\Delta U_t
\end{bmatrix}\right)
\begin{bmatrix}
	\Delta X_t\\
	\Delta U_t
\end{bmatrix}^{\dagger},
\end{align}
for all $t=0, \ldots, T-1$.

\paragraph*{ II) Continuous differentiability of $[\Delta A_t, \Delta B_t]$ in \eqref{eq:expression_for_delta_AB}}

Notice that $o_\traj(\cdot)$ in \eqref{eq:expression_for_delta_AB} must be a continuously differentiable function of $\traj, \Delta X_t, \Delta U_t$ by Assumption \ref{hp:f_ell_regularity} (since it is a stack of the $o_\traj(\cdot)$ in \eqref{eq:first_approx}, which in turn must be $C^1$ in their argument being the remainder of a first-order approximation of a $C^2$ function). Furthermore, if $(\Delta X_t, \Delta U_t)\in \mathcal{F}_M$ for all $t$, then also the pseudoinverse in \eqref{eq:expression_for_delta_AB} is a $C^1$ function of the the entries of $\Delta X_t, \Delta U_t$, since it can be calculated as
\begin{equation}\label{eq:pseudoinverse}
\begin{bmatrix}
	\Delta X_t\\
	\Delta U_t
\end{bmatrix}^{\dagger} = \begin{bmatrix}
	\Delta X_t\\
	\Delta U_t
\end{bmatrix}^{\top} 
\left(
\begin{bmatrix}
	\Delta X_t\\
	\Delta U_t
\end{bmatrix}
\begin{bmatrix}
	\Delta X_t\\
	\Delta U_t
\end{bmatrix}^{\top}
\right)^{-1},
\end{equation}
and this is a product of $C^1$ functions of the entries of $\Delta X_t, \Delta U_t$ in the considered domain.
This proves that, given $(\Delta X_t, \Delta U_t)\in \mathcal{F}_M$ , for all $t=1, \ldots, T-1$ the error matrix $[\Delta A_t, \Delta B_t]$ in \eqref{eq:expression_for_delta_AB} is a $C^1$ function of $\Delta X_t, \Delta U_t$.
In turn, this implies that the function $\Delta_{AB}(\traj, \bDx, \bDu)$ is a $C^1$ function of $\bDx, \bDu$, since it is the stack of all $[\Delta A_t, \Delta B_t]$. 

\paragraph*{III) Local linear bound on $\|\Delta_{AB}\|$}

Denote, for simplicity, 
\begin{equation}
\Delta_t\coloneqq \begin{bmatrix}
	\Delta X_t\\
	\Delta U_t
\end{bmatrix}.
\end{equation}
By substituting \eqref{eq:pseudoinverse} in \eqref{eq:expression_for_delta_AB}, we obtain
\begin{equation}
\begin{split}
	\begin{bmatrix}
		\Delta A_t & \Delta B_t
	\end{bmatrix}	
	& = o_\traj(\Delta_t)\Delta_t^\top (\Delta_t \Delta_t^\top)^{-1}\\
	& = o_\traj(\Delta_t\Delta_t^\top)(\Delta_t \Delta_t^\top)^{-1},
\end{split}
\end{equation}
and by taking the norms and using Assumption \ref{hp:well_posed},
\begin{equation}
\begin{split}
	\left\|
	\!\begin{bmatrix}
		\Delta A_t & \Delta B_t
	\end{bmatrix}\!\right\|
	&\! \leq \! \|o_\traj(\Delta_t\Delta_t^\top)\| \|(\Delta_t \Delta_t^\top)^{-1}\|\\
	&\! \leq \! \|o_\traj(\Delta_t\Delta_t^\top)\| \|\Delta_t \Delta_t^\top\|^{-1}\kappa(\Delta_t\Delta_t^\top)\\
	&\! \leq \! \|o_\traj(\Delta_t\Delta_t^\top)\| \|\Delta_t \Delta_t^\top\|^{-1}M
\end{split}
\end{equation}
Notice that by definition, $o_\traj(\cdot)$ goes to zero faster than its argument, i.e., for any $\epsilon>0$ we can find $\delta>0$ such that, for all $t=0, \ldots, T-1$, it holds
\begin{equation}\label{eq:little_o}
\| \Delta_t \Delta_t^\top \|\leq \delta \implies 
\|o(\Delta_t \Delta_t^\top)\|
\|\Delta_t \Delta_t^\top\|^{-1} \leq \epsilon.
\end{equation}	
Being $[\Delta A_t\;\Delta B_t]$ continuously differentiable in $\traj, \Delta X_t, \Delta U_t$ in the considered domain, we know that for any compact set $\mathcal{H}\times\mathcal{K} \subset\Tr \times \mathcal{F}_M$, for all fixed $\traj \in \mathcal{H}$ and $\Delta_t \coloneqq (\Delta X_t, \Delta U_t)$, $\Delta^\prime_t\coloneqq (\Delta X_t^\prime, \Delta U^\prime_t)\in \mathcal{K}$ there exist $g(\mathcal{H}, \mathcal{K})>0$ such that
\begin{equation}
\begin{split}
	&\left\|\begin{bmatrix}
		\Delta A_t & \Delta B_t
	\end{bmatrix}
	-
	\begin{bmatrix}
		\Delta A^\prime_t & \Delta B^\prime_t
	\end{bmatrix}\right\|
	\leq g(\mathcal{H}, \mathcal{K}) \left\| 
	\Delta_t-
	\Delta_t^\prime
	\right\|\\
	&\left\|\!
	\begin{bmatrix}
		\Delta A_t & \Delta B_t
	\end{bmatrix}\!
	\right\|
	\!\leq\!  \left\|\!
	\begin{bmatrix}
		\Delta A^\prime_t & \Delta B^\prime_t
	\end{bmatrix}\!\right\| \!+\! g(\mathcal{H}, \mathcal{K})( \| \Delta_t\| \!+\!\|\Delta_t^\prime\|)\\
	\vphantom{\int}
\end{split}
\end{equation}
Using \eqref{eq:little_o}, given any $\epsilon>0$, there exists $\delta>0$ for which, by picking $\|\Delta^\prime\|\leq \delta$, we can ensure $\|[\Delta A_t^\prime \; \Delta B_t^\prime]\|\leq \epsilon M$, from which
\begin{equation}
\begin{split}
	\left\|\!
	\begin{bmatrix}
		\Delta A_t & \Delta B_t
	\end{bmatrix}
	\right\|
	\leq &  g(\mathcal{H}, \mathcal{K}) \|\Delta_t\| + g(\mathcal{H}, \mathcal{K})\delta + \epsilon M.
\end{split}
\end{equation}
Since $\epsilon$ can be picked arbitrarily small, we obtain it must hold
\begin{equation}
\begin{split}
	\left\|\!
	\begin{bmatrix}
		\Delta A_t & \Delta B_t
	\end{bmatrix}\!
	\right\|
	\leq &  g(\mathcal{H}, \mathcal{K}) \|\Delta_t\|,
\end{split}
\end{equation}
and since this holds for all $t=0, \ldots, T-1$, we have
\begin{equation}
\begin{split}
	\|\Delta_{AB}(\traj, \bDx, \bDu)\| 
	\leq &  g(\mathcal{H}, \mathcal{K}) T \|( \bDx, \bDu)\|
\end{split}
\end{equation}
for all $(\traj, \bDx, \bDu)\in \mathcal{H}\times \mathcal{K}$.

\subsection{Proof of Lemma~\ref{lemma:solution_distances}}
\label{proof:lemma:solution_distances}

Consider the exact problem~\eqref{eq:pronto_descent}. We reformulate it as 
\begin{align}\label{prob:real}
\min_{\zeta\in \R^s} \;\;\; &  \zeta^\top \dd^2 \ell(\traj) \zeta + \dd \ell(\traj)^\top\zeta\\
\text{s.t.} \;\;\; & H(\traj)\zeta = 0, \nonumber
\end{align}
with $\zeta = (\bdeltax, \bdeltau)$ and
\begin{equation}\label{eq:H}
\begin{split}
	H(\traj) &\coloneqq
	\begin{bmatrix}
		H_x(\traj) & H_u(\traj)
	\end{bmatrix}, \\
	H_x(\traj) &:= 
	\begin{bmatrix}
		I 			& 0 & 0& 0 \\
		-A_0 	& I 		&0 & 0 \\
		0	& \ddots  & \ddots& 0\\
		0	& 0  & -A_{T-1}& I 
	\end{bmatrix} \in \R^{s_x\times s_x}\\
	H_u(\traj) &:= 
	\begin{bmatrix}
		0& 0& 0\\
		-B_0 &0 &0 \\
		0& \ddots& 0\\
		0& 0&-B_{T-1} 
	\end{bmatrix}\in \R^{s_x\times s_u}.
\end{split}
\end{equation}
The estimated problem \eqref{prob:pronto_approx} differs from this only in the constraints, i.e., it can be written as
\begin{align}\label{prob:approx}
\min_{\zeta\in \R^s} \;\;\; &  \zeta^\top \dd^2 \ell(\traj) \zeta + \dd \ell(\traj)^\top\zeta\\
\text{s.t.} \;\;\; & \hat{H}(\traj, \bDA, \bDB)\zeta = 0, \nonumber
\end{align}
where $\hat{H}(\traj, \bDA, \bDB) = H(\traj)+ \tilde{H}(\bDA, \bDB)$ given by
\begin{equation}\label{eq:tildeH}
\begin{split}
	\tilde{H}(\bDA, \bDB)&:=
	\begin{bmatrix}
		\tilde{H}_x(\bDA, \bDB) &\tilde{H}_u(\bDA, \bDB)
	\end{bmatrix}, \\
	\tilde{H}_x(\bDA, \bDB) &\coloneqq
	\begin{bmatrix}
		0		& 0 & 0& 0 \\
		-\Delta A_0 	& 0 		&0 & 0 \\
		0	& \ddots  & \ddots& 0\\
		0	& 0  & -\Delta A_{T-1}& 0 
	\end{bmatrix} \in \R^{s_x\times s_x}\\
	\tilde{H}_u(\bDA, \bDB) &\coloneqq
	\begin{bmatrix}
		0& 0& 0\\
		-\Delta B_0 &0 &0 \\
		0& \ddots& 0\\
		0& 0&-\Delta B_{T-1} 
	\end{bmatrix}\in \R^{s_x\times s_u}.
\end{split}
\end{equation}
By Assumption \eqref{hp:newton}, \eqref{prob:real} is a strictly convex program and has a unique minimizer $\zeta^\star(\traj)$. Furthermore, given the structure of $H(\traj)$, Linear Independence Constraint Qualification (LICQ) holds for any $\traj \in \R^s$. This means that KKT conditions hold for the solution $\zeta^\star(\traj)$ of problem \eqref{prob:real} \cite[Thm. 3.14]{diehl2011numerical}. By \cite[Thm. 3.18]{diehl2011numerical}, having only equality constraints, also SOSC holds for $\zeta^\star(\traj)$. We can apply \cite[Thm. 3.19]{diehl2011numerical} and obtain that there exists a $\delta_{AB}(\traj)>0$ such that, for all $\|\tilde{H}(\bDA, \bDB)\|\leq \delta_{AB}(\traj)$, there exists a unique solution $\hat{\zeta}^\star(\traj, \bDA, \bDB)$ to \eqref{prob:approx} and it depends differentiably on $\tilde{H}(\traj, \bDA, \bDB)$. 
Furthermore, $\delta_{AB}(\traj)$ depends continuously on $\traj$, since $\dd^2\ell(\traj), \dd \ell(\traj), H(\traj)$ describing the QP \eqref{prob:real} are all continuous in $\traj$ by Assumption \ref{hp:f_ell_regularity}.

Next, as $\tilde{H}(\bDA, \bDB)$ is linear in $\traj, \bDA, \bDB$, also $\hat{\zeta}^\star(\traj, \bDA, \bDB)$ is differentiable in $\traj, \bDA, \bDB$. This means that $\zeta^\star(\traj)-\hat{\zeta}^\star(\traj, \bDA, \bDB)$ is smooth in its arguments.

To conclude the statement, notice that if $\bDA, \bDB=0$ then $\tilde{H}(\bDA, \bDB)=0$, and thus, since problems \eqref{prob:real} and \eqref{prob:approx} become identical, $\zeta^\star(\traj)-\hat{\zeta}^\star(\traj, 0, 0)=0$.

\bibliography{biblio_data-driven_pronto}
\bibliographystyle{IEEEtran}

\vfill\null

\end{document}